\documentclass[12pt]{article}
\usepackage{amsfonts}

\makeatletter
\@addtoreset{equation}{section} 
\makeatother

\newcommand{\eqn}[1]{(\ref{#1})}

\def\appendix#1
{
\addtocounter{section}{1}\setcounter{equation}{0}
 \renewcommand{\thesection}{\Alph{section}}
 \section*{Appendix~\thesection\protect\indent \parbox[t]{11.715cm}{#1}}
 \addcontentsline{toc}{section}{Appendix \thesection\ \ \ #1}
}
\newcommand{\complex}{{\mathbb{C}}} 
\newcommand{\zed}{{\mathbb{Z}}} 
\newcommand{\real}{{\mathbb{R}}} 
\newcommand{\reals}{{\mathbb{R}}} 
\newcommand{\realb}{{\mathbb{R}}} 
\newcommand{\torus}{{\mathbb{T}}} 
\newcommand{\id}{{\mathbb{I}}} 
\newcommand{\ket}[1]{\left|#1\right\rangle}

\def\<#1,#2>{\left\langle#1,#2\right\rangle} 

\def\hil{\mathcal{H}}

\def\nn{\nonumber}
\newcommand\opname[1]{\mathop{\mathrm{#1}}\nolimits}

\newcommand{\Tr}{\opname{Tr}}
\newcommand{\wres}{\opname{wres}} 
\def\e{{\,\rm e}\,}

\def\be{\begin{equation}}
\def\ee{\end{equation}}
\def\beq{\begin{eqnarray}}
\def\eeq{\end{eqnarray}}
\def\bd{\begin{displaymath}}
\def\ed{\end{displaymath}}

\newcommand{\fr}[2]{{\textstyle\frac{#1}{#2}}}
\newcommand{\half}{{\textstyle\frac{1}{2}}}

\newcommand{\bea}{\begin{array}}
\newcommand{\ea}{\end{array}}

\newcommand{\sg}{\sigma}

\newcommand{\del}{\partial}

\newcommand{\eps}{\epsilon}

\newcommand{\g}{{\mathcal{G}}}

\newcommand{\A}{{\mathcal{A}}}

\begin{document}
\begin{titlepage}
\begin{flushright}
\baselineskip=12pt
DSF--40--01\\
BiBoS 01-12-068\\
ESI 1109\\
hep--th/0112092
\hfill{ }\\
December 2001

\end{flushright}

\begin{center}

\baselineskip=24pt

{\Large\bf Infinitely many star products to play with}

\baselineskip=14pt

\vspace{1cm}

{\bf J.M.\ Gracia-Bond\'{\i}a} $^{a,b}$, {\bf F.\ Lizzi} $^{b}$, {\bf G.~Marmo}
$^{b}$ and {\bf P. Vitale} $^{c}$\\[6mm]
$^a$ {\it BiBoS, Fakult\"{a}t der Physik, Universit\"{a}t Bielefeld, 33615
Bielefeld, Germany}\\[6mm]
$^b$ {\it Dipartimento di Scienze Fisiche, Universit\`{a} di Napoli {\sl
Federico II}\\
and {\it INFN, Sezione di Napoli, Monte S.~Angelo}\\
Via Cintia, 80126 Napoli, Italy}\\
{\tt fedele.lizzi@na.infn.it, giuseppe.marmo@na.infn.it}\\[6mm]
$^c$ {\it Dipartimento di Fisica, Universit\`{a} di Salerno and INFN
\\ Gruppo Collegato di Salerno, Via S. Allende
\\84081 Baronissi (SA), Italy}
\\ {\tt patrizia.vitale@sa.infn.it}\\

\end{center}

\vskip 2 cm

\begin{abstract}

While there has been growing interest for noncommutative
spaces in recent times, most examples have been based on the
simplest noncommutative algebra: $[x_i,x_j]=i\theta_{ij}$.
Here we present new classes of (non-formal) deformed
products associated to linear Lie algebras of the kind
$[x_i,x_j]=ic_{ij}^kx_k$. For all possible three-dimensional
cases, we define a new star product and discuss its
properties. To complete the analysis of these novel
noncommutative spaces, we introduce {\em noncompact\/}
spectral triples, and the concept of {\em star triple}, a
specialization of the spectral triple to deformations of the
algebra of functions on a noncompact manifold. We examine
the generalization to the noncompact case of Connes'
conditions for noncommutative spin geometries, and, in the
framework of the new star products, we exhibit some
candidates for a Dirac operator. On the technical level,
properties of the Moyal multiplier algebra
$M({\real}_\theta^{2n})$ are elucidated.

\end{abstract}

\end{titlepage}

\section{Introduction}

Over five years ago, Connes gave the first axiomatics for first-quantized
fermion fields on (compact) noncommutative varieties, the so-called
spectral triples~\cite{ConnesGravplus}. Shortly afterwards, it was realized
that compactification of matrix models in M-theory leads to noncommutative
tori~\cite{CoDoS}. The next logical step, introducing quantum fields in the
framework of Connes' noncommutative spaces, was taken in~\cite{Atlas}. Then
it came the discovery by Seiberg and Witten that the dynamics of open
strings, ``rigidified" by the presence of a magnetic field, is described by
a noncommutative geometry associated to the Moyal product~\cite{SMonster}.
Since then, field theories on Moyal-type spaces, including noncommutative
tori, have been scrutinized extensively; good reviews
are~\cite{DouglasN,Szabo}.

The magnetic field breaks of course Lorentz invariance; a model which
emphasizes full Poincar\`e invariance had been proposed in~\cite{DFR}.

By now it is clear that (as it was bound to happen) Moyal-type
spaces do {\it not\/} suffice to the needs of string and
noncommutative field theories. Apart from the half-string
product introduced long ago by Witten to construct a string
field theory \cite{WittenSFT}, there has been the appearance
of ``fuzzy sphere" products in the theory of strings and
branes \cite{fuzzystr}. Moreover, the nonlocality of
noncommutative gauge theories induces, at the level of the
effective actions, generalizations of the Moyal product,
including a ternary product~\cite{genprod}. These new
products are most often either products of finite matrices
(the fuzzy algebras) or can be reduced to the Moyal product
---also for the Witten product~\cite{Bars}.
On the formal side, a more ambitious attempt has been made
in~\cite{MonacoGang}.

Very recently, there has been an effort~\cite{HLS} to face
the shortfall, by introducing a non-formal star product
on~$\real^3$. (The Moyal product on $\real^{3}$ is trivial
in the sense that it just extends the Moyal product on
$\real^{2}$, with one direction remaining ``totally
commutative".) The construction relies on ``projecting" to
the latter, with the help of the Hopf fibration, a relative
of the Moyal product on $\real^4$: to wit, the twisted
product associated to normal ordering. In other words, it
harks back to the Jordan-Schwinger map. The product
in~\cite{HLS} is rather directly related to the fuzzy
sphere.

The method by Hammou and coworkers in~\cite{HLS} is too
dependent on specifics to be readily generalized. Perhaps
the time has come to {\it upgrade\/} the fabrication of star
products to the industrial stage. This is one theme of the
present paper.

To construct viable generalizations it is required, at the outset, to have
noncommutative geometries which are analytically controlled
---as the Moyal product is. That is to say, formal deformations of
classical manifolds, of which there are plenty~\cite{Maxim}, do not
nearly foot the bill. We will explicitly construct analytically
controlled products based on each and every three dimensional Lie
algebra; as the Moyal product generalizes the Heisenberg relation
$[x_i,x_j]=i\theta_{ij}$, our deformed algebras are based on the
relations $[x_i,x_j]=ic^k_{ij} x_k$, with constant $c$'s.

Meanwhile, in a more mathematical vein, new \textit{compact\/} spectral
triples, i.e., noncommutative compact spin geometries, have been
introduced in~\cite{ConnesLa}. Namely, Connes and Landi there explained
what an even-dimensional noncommutative sphere is. Related to this,
V\'{a}rilly introduced noncommutative orthogonal groups~\cite{Larissa}.
See also~\cite{dobleonada} in this connection.
Connes and Dubois-Violette~\cite{ConnesViolette} have dealt in depth with the
odd-dimensional case and, more recently, Figueroa, Landi and
V\'{a}rilly~\cite{Elara} extend the constructions to cover Grassmannians
and other homogeneous spaces as well.%
\footnote{
Let us point out here that, in the continuous category, noncommutative
spheres were first introduced by Matsumoto~\cite{Banzai}, and that 
some distinctions of principle need to be drawn between Connes--Landi and
Connes--Dubois-Violette spheres; also,
the question of eventual symmetries of the latter
from deformations of orthogonal groups seems to be open.}

The deformation, in the sense of Connes--Dubois-Violette, of a
sphere $\mathbb{S}^n$ is obtained via an algebraic deformation of the
noncompact manifold $\real^{n+1}$, plus a reduction to a quotient algebra
by setting $r^2=1$, where $r^2$ is a central quadratic element in the
deformed $\real^{n+1}$ algebra. The method would not seem to work on
$\real^{3}$: the only spherical spectral triple in two dimensions is the
ordinary sphere.

The definition of spectral triple, giving rise to noncommutative spin
manifolds, in~\cite{ConnesGravplus} had several strong restrictions and
mathematical conditions (see Section~6 below) destined to ensure that
the commutative case realizes a fully algebraic description of a compact
spin manifold. The framework needs enlargement, and in fact, postulates
for Riemannian, not necessarily spin, manifolds~\cite{Lord} and for
semi-Riemannian (for instance, with Lorentzian signature) spectral
triples~\cite{MainzplusLeipzig} have been proposed. However, in all of
these generalizations, compactness is retained. This is a drawback from
the standpoint of noncommutative field theory.

As its second theme, and partly with the aim of fitting the new star
products into a general analytic framework, this paper introduces {\it
noncompact\/} noncommutative geometries. Going into this new territory, the
landscape changes, and the relation of the different natural examples
becomes intricate. It turns out that, {\it mutatis mutandis\/}, the whole
``seven-axiom" apparatus by Connes, can be pushed through to the noncompact
case. This covers the commutative bedrock and, we conjecture, the
noncompact spaces associated to the Connes--Landi and
Connes--Dubois-Violette spheres, insofar as they are spectral triples.
However, the Moyal algebra case is {\it not\/} covered by the more direct
generalization.

The difficulties cluster around the ``dimension axioms", numbered 1 and 6
in our reckoning (see Section~6). In the noncompact situation, two
definitions of ``classical dimension" part ways. These are: (i) the
``metric dimension", based on the asymptotic growth of the spectrum of the
resolvent of the Dirac operator $D$, and (ii) the ``homological dimension",
based on the Hochschild homology of the underlying algebra, which a priori
has nothing to do with $D$. The Hochschild homology of the Moyal algebra
vanishes in degrees greater than zero; and so in some sense the dimension
of that noncommutative space is zero; this will be confirmed by the Dixmier
trace check.

At the present stage of exploration of the vast
noncommutative world, it would seem nevertheless unwise to
rule out the Moyal product, which centrality is rather
enhanced by the first part of our investigation, for the
sake of axiomatics.

So, by relaxing the axioms, we introduce as well a second
class of noncompact geo\-metries, that we term {\it star
triples}. A star triple is a kind of spectral triple in
which: a) the noncommutative algebra is obtained from a
deformation, i.e., it is given by a star product on a linear
space of functions and/or distributions on a given manifold;
b) the Dirac operator (is possibly deformed, but) remains an
ordinary (pseudo-)differential operator on that original
manifold; this can be used to establish the dimension.
Almost all known spectral triples are of this type. The star
triple set of postulates recovers the Moyal subcase.

Let us return to our first subject. The strategy for that can be summarized
in a few words: we look for {\it subalgebras\/} of the Moyal algebra, that
realize, in the spirit of Dirac, deformations of Poisson structures on
$\real^n$. Before indicating how some Moyal subalgebras give rise to the
star products, let us point out that the paradigm of noncommutative
manifolds, the venerable noncommutative
torus~\cite{ConnesTorusplusRieffelRot}, can be as well rigorously proven to
be, after all, a Moyal subalgebra. See our discussion in Section~5.

Our approach asks for familiarity with Moyal algebra. When
Seiberg and Witten related the dynamics of strings to the
Moyal product, they were not entering mathematically virgin
land. For more than fifty years, going back
to~\cite{GroeJoseE}, with the precedent of~\cite{Wigner},
that product has been used to perform quantum mechanical
calculations on phase space. Among the tools we import to
noncommutative field theory from quantum mechanics in phase
space, the Wigner transform~\eqn{WignerTr} does not seem to
have been used before. We have collected some of these in a
long appendix, for the convenience of the reader. We
consider that it pays to be acquainted with the body of
extant results in the literature of phase-space quantum
mechanics. Otherwise, one risks to miss the more effective
approaches, costly mistakes (more on that in Section~6 and
the appendix) or, in the best of cases, laboured arguments
to rediscover trivialities.

\smallskip

The approaches used in recent constructions of noncommutative geometries in
the sense of Connes~\cite{ConnesLa,Larissa,ConnesViolette,Elara} have in
common that noncommutative tori are exploited to the full: the basic idea
is to find actions of torus groups ${\cal T}^l$ (where $l\geq2$) on the
spheres and the orthogonal groups, and to twist them. This way
noncommutativity is ``injected" in the manifold from within, so to speak.
Here, as advertised, we take a different tack: to ``project" it, in the
noncompact case, from known noncommutative manifolds ---in the occurrence,
the standard Moyal algebras.

To carry on our programme, suitable maps
$\pi:\,\real^{2n}\to\real^{d}$ are needed; a Poisson
structure $\{.,.\}_P$ on $\real^d$ is assumed given. In
fact, such maps $\pi$ are the oldest game in town. They were
introduced by Lie, under the name of {\it
Funktionengruppen}, or ``function groups"
---see~\cite{SophusI} and also~\cite{BeppeofoldWeinstein}. A
Funktionengruppe in the sense of Lie is a collection $\cal F$ of functions
of the canonical variables $(q_i,p_j)$ on $\real^{2n}$ such that:

\begin{itemize}
\item $\cal F$ is closed by functional composition, and generated by a
finite number (say, $d$) of its elements.
\item $\cal F$ closes to a subalgebra of $C^\infty({\real}^{2n})$ under the
Poisson bracket.
\end{itemize}

In local terms, the construction of Funktionengruppen is essentially
equivalent to the problem of finding Poisson maps $\pi$ from $\real^{2n}$
to (suitable subsets of) $\real^{d}$. Those we can call (linear) symplectic
realizations of the Poisson structure on the latter. In fact, ${\cal F}:=
\pi^*C^\infty({\real}^d)$ gives the function group.%
\footnote{Since the maps $\pi^*$ are different for the several
function groups considered in this paper, we should actually use the
notation $\pi^*_{\cal G}$; we will however omit the subscript not to
burden the notation.} But the original formulation by Lie is more in the
spirit of noncommutative geometry.

This paper revolves around the observation that, under favourable
circumstances, Lie's Funktionengruppen are also closed under the Moyal
product.

For simplicity, in this connection, we concentrate on nontrivial
deformations of $\real^{3}$ that can be derived from Moyal products on
$\real^{4}$. Deformations of $\real^{3}$ are of interest for D3-brane field
theories in which the brane's time direction is commuting; these are the
relevant ones for the AdS/CFT correspondence. Also, they can be applied to
world-volumes of Euclidean D2-branes. Our scheme, nevertheless, does not
suffer any limitation by dimension, and its wider applicability will be
evident.

We begin, in Section~2, by classifying all Poisson
structures on~$\real^3$. There is a wealth of them. A
particularly attractive class of Funktionengruppen related
to linear Poisson brackets is selected in Section~3. The
reader is advised to at least scan the appendix for the
notation, before going further. Section~4 is devoted to the
proof of the main theorem. The corresponding star products
are constructed in Section~5. Then we examine properties of
the resulting algebras case by case. Section~6 turns to
noncompact noncommutative spin geometries and star triples.
We indicate how a direct extension of Connes'axioms covers
the basic (compact and) commutative noncompact cases, and
also show that Moyal algebras are noncompact star triples in
our sense. In Section~7, we examine the standing of the new
star products in regard to noncompact star triples
postulates. Finally, in Section~8, we summarize the results,
and indicate some of the promising avenues thereby open.

\section{Poisson structures on~$\realb^3$}

We indicate the coordinates on~$\real^4$ with a ``phase space'' notation:
$\real^4\ni u=(q_1,q_2,p_1,p_2)$. To
proceed, we review the classification procedure of Poisson structures on
$\real^3$, due to Grabowski, Perelomov and one of us~\cite{marpergrab}.

The basic idea, to classify bivector fields $\Lambda$ that
could lead to Poisson structures on orientable manifolds, is
to effect the inner product of $\Lambda$ with the
fundamental form. This allows to formulate (among other
things) integrability conditions in terms of forms. On
$\real^3$ it leads to a Casimir 1-form.

Consider coordinates provisionally called $(x_1, x_2, x_3)$ on~$\real^3$,
and the fundamental form $\Omega = dx_1 \wedge dx_2 \wedge dx_3$. Let
$\Lambda$ denote the bivector field corresponding to any given Poisson
bracket:
\be
\Lambda = \sum_{i<j}c_{ij}(x)\frac{\del}{\del x_i}\wedge\frac{\del}{\del x_j};
\ee
or intrinsically $\Lambda(df \wedge dg)=\{f,g\}$. By contracting $\Lambda$
with $\Omega$ we find
\be
i_\Lambda \Omega = \half \epsilon_{ijk} c_{ij}(x) dx_k
\ee
i.e.,
\be
i_\Lambda \Omega = A_k(x) dx_k =: \alpha,
\ee
with $A_k = {1\over 2}\epsilon_{ijk}c_{ij}(x)$. The Jacobi identity for the
Poisson brackets associated to $\Lambda$
$$
\{x_i,x_j\} = c_{ij}(x)
$$
is easily seen to be equivalent to the integrability condition for the
Pfaff equation associated to $\alpha$:
\be
d\alpha \wedge \alpha = 0 \label{alphadalpha},
\ee
that is to say, in classical language
\be
\vec A {\cdot} \opname{curl} \vec A = 0.
\ee
In conclusion, Poisson structures are characterized by 1-forms which admit
an integrating factor; this characterization turns to be useful for our
purposes. Locally we have $\alpha=f d\phi$ and
$$
\{x_i,x_j\} = \epsilon_{ijk} f \frac {\del \phi}{\del x_k}.
$$
The symplectic leaves for $\Lambda$ are the level sets of $\phi$, on which
(the pull--back of) $\alpha$ vanishes. The 1-form $\alpha$ is what we call
the {\it Casimir form\/} for the given brackets.

{}From now on, we consider a restricted class of Poisson structures, for
which the components of the bivector field $\Lambda$ are linear in the
coordinates (in particular we reluctantly refrain here from studying the
quadratic case, related to Sklyanin-type algebras). That is, $\alpha$ has
the particularly simple form $A_j(x) = M_{ij}x_i$, with $M$ a constant
matrix. This is necessary and sufficient to describe all the three
dimensional Lie algebras over the reals.

A real finite-dimensional Lie algebra ${\cal G}$ with Lie
bracket $[{\cdot},{\cdot}]$ defines in a natural way a
Poisson structure $\{{\cdot},{\cdot}\}$, on the dual space
${\cal G}^*$ of ${\cal G}$. One is allowed to think of
${\cal G}$ as a subset of the ring of smooth functions
$C^\infty ({\cal G}^*)$. Choosing a linear basis
$\{E_i\}_1^n$ of ${\cal G}$, and identifying them with
linear coordinate functions $x_i$ on ${\cal G}^*$ by means
of $x_i(x) = \<x,E_i>$ for all $x\in {\cal G}^*$, we define
the fundamental brackets on ${\cal G}^*$ by the expression
$\{x_i,x_j\}_{{\cal G}}=c_{ij}^k x_k$ where
$[E_i,E_j]=c_{ij}^k E_k$ and $c_{ij}^k$ denote the structure
constants of the Lie algebra. The Poisson bracket
$\{.,.\}_{{\cal G}}$ is associated to a bi-vector field
$\Lambda$ which is locally given by
\be
\Lambda = \sum_{i<j}c_{ij}^k x_k {\del\over\del x_i}\wedge {\del\over\del x_j}.
\ee
Here the $c_{ij}^k := \epsilon_{ijl}M_{kl}$ have all the required
properties.

The classification of three dimensional Lie algebras goes back to
Bianchi~\cite{Bianchi}; see also~\cite{Jacobson}. We present here such
classification, which will give rise to inequivalent star products on
$\real^3$, in terms of the one-form
$\alpha=M_{ij}x_idx_j$~\cite{CIMP,MMVZ}.

Decompose $\alpha$ into its symmetric and antisymmetric parts:
\be
\alpha= \half(M_{ij} - M_{ji}) x_i\,dx_j + \half(M_{ij} +
M_{ji}) x_i\,dx_j. \label{alpham}
\ee
Defining $a_k=\epsilon_{ijk} M_{ij}$ we get
\be
\alpha= \half a_k\epsilon_{ijk}(x_i\,dx_j-x_j\,dx_i) +\half\,d(M_{ij}
x_i x_j).
\ee
Recall that the Jacobi identity implies $d\alpha\wedge \alpha=0$. We have
two cases: either the form $\alpha$ is closed (and therefore exact
since we are in $\real^3$):
\be
\alpha= \half\,d\,(M_{ij} x_i x_j) ~~~~~~    i,j\in\{1,2,3\}.\label{closed}
\ee
Or there exists a vector field, say $X_1={\del \over \del x_1}$ , which is
in the
kernel of both $\alpha$ and $d\alpha$. In this case we have
\be
\alpha= h(x_2\,dx_3-x_3\,dx_2) + \half\,d(M_{ab}x_a x_b) ~~~~a,b\in\{2,3\},
\label{notclosed}
\ee
with the constant $h\ne 0$. We can bring both cases to
normal form. The first by diagonalizing the symmetric part
of $M$, while for the second we use a rotation matrix which
preserves the form $x_2\,dx_3-x_3\,dx_2$. We find therefore,
using the notation $x=x_1, y=x_2$ and $w=x_3$:

\begin{itemize}

\item[A.] For the case of $\alpha$ closed:
$$
\alpha= \half\,d(a x^2+b y^2+cw^2).
$$
\item[B.] For the case of $\alpha$ not closed:
$$
\alpha= \half  h(y\,dw-w\,dy) + \half\,d (b y^2+c w^2).
$$

\end{itemize}

In summary, any three dimensional Lie algebra is characterized by the Casimir
form
\be
\alpha = h(y\,dw-w\,dy) +  \half\, d (ax^2+by^2+cw^2)  \label{alph}
\ee
when the real parameters $h,a,b,c$ are appropriately selected. This yields
the Poisson brackets
\be
\{x,y\}=cw+hy,~~~~
\{y,w\}=ax,~~~~
\{w,x\}=by-hw \label{classbr}
\ee
while the Jacobi identity
\be
d\alpha\wedge \alpha = 2h\,a\,x\;dy\wedge dw\wedge dx = 0
\ee
holds true if and only if $ha=0$. Thus we have two essentially different
classes of algebras: those corresponding to the closed Casimir form ($h=0$,
case A), and ($a=0$, case B), those corresponding to the Casimir
form~\eqn{notclosed}.

We list the algebras, indicating their relation to Bianchi's~\cite{Bianchi}
classification of three dimensional Lie algebras in nine families,
according to the dimension of ${\cal G}'$, the derived algebra. Our
approach gives rise to ten families.

In the first situation, when $\alpha=d{\cal C}$, let us call ${\cal C}$ a
Casimir function. In case A the parameters $a,b,c$, when different from
zero, may be all normalized to modulus one. This grouping includes six
different isomorphism classes of Lie algebras:

\begin{itemize}
\item[A.1] $su(2)\simeq so(3)$ with $a,b,c$ all different from 0 and of the
same sign. A basis can be chosen so that $a=b=c=1$ (similar remarks will be
understood in what follows, when pertinent). This corresponds to type~$IX$
from Bianchi's classification.
\item[A.2] $e(2)$, the algebra of the Euclidean group in two dimensions, which
may be obtained by contraction from the previous class, say $a\to 0$. This
corresponds to an algebra of type~$VII$ in Bianchi's classification, that
we can term~$VII_0$.
\item[A.3] $sl(2,\real)\simeq su(1,1)\simeq so(2,1)$, with $a,b,c$ all
different from 0 and of different sign. This corresponds to type~$VIII$.
\item[A.4] $iso(1,1)$, the Poincar\'{e} algebra in two dimensions, which may be
obtained by contraction from the previous algebra. This corresponds
to a type $VI$ algebra; let us call it~$VI_0$.
\item[A.5] $h(1)$, the Heisenberg-Weyl algebra, with only one parameter
different
from zero, for example $c> 0$. It may be obtained by further contraction
from both $e(2)$ and $iso(1,1)$. This is type~$II$.
\item[A.6] The abelian algebra with a=b=c=0. This corresponds to type~$I$.
\end{itemize}

The second case, with $\alpha$ not exact, includes four families of Lie
algebras, two of which are further subclassified by the real parameter
$bc/h^2$, the transformations of $\alpha$ leaving invariant the ratio
between the determinants of the symmetric and antisymmetric parts of the
matrix $M_{ij}$. We recall that all algebras of type B have $a=0$ because
of the Jacobi identity. One has
\begin{itemize}
\item[B.1] $h=1,b=c=0$, that is $sb(2,\complex)$, the Lie algebra of the
group of
$2{\times}2$ upper(lower) triangular complex matrices with unit
determinant. This
corresponds to type~$V$.
\item[B.2] $h=1,b=0,c=1$; this yields Bianchi's~$IV$.
\item[B.3] $h\ne0,b=1,c=-1$. When $h=1$, then $\dim{\cal G}'=1$ and this does
correspond to type $III$. All the others, for which $\dim{\cal G}'=2$, are
type~$VI$; we denote them by~$VI_h$.
\item[B.4] $h\ne0,b=c=1$: this is type~$VII_h$.
\end{itemize}
Cases B.3 and B.4 are one-parameter families, as it is impossible to put
all parameters equal to one with a similarity transformation. We will refer
collectively to the previous four classes of algebras as ${\cal G}_h$.
Notice that the Bianchi classification mixes algebras with closed and not
closed Casimir form. Indeed, types~$VI_0$ and~$VII_0$ correspond to closed
one-forms, while~$VI_h$ and~$VII_h$ with $h\ne 0$ do not. Our
classification exhausts all possible three dimensional algebras.

\section{Funktionengruppen}

Consider now $\real^4$ with the canonical symplectic structure given by the
Poisson brackets
$$
\{q_i,p_j\} = \delta_{ij},
$$
associated to the symplectic form
$$
\omega = dq_1\wedge dp_1 + dq_2\wedge dp_2.
$$
We are ready to show realizations
$\pi:\real^4\rightarrow\g^*\equiv\real^3$. We express $\pi$
through the change of variables $\pi^*$ that pulls smooth
functions on $\real^3$ back to smooth functions on
$\real^4$; in other words, we give the realizations by means
of contravariant arrows in the category of smooth functions
and maps. All that one has to do is to find three
independent functions $f_1,f_2,f_3$ on $\real^4$ whose
corresponding canonical brackets\footnote{The corresponding
brackets on $\real^3$ are Kostant--Kirillov--Souriau
brackets, and thus the images of the realizations are unions
of orbits of the coadjoint action; this theory is well known
and does not bear repetition here.} have the required
form~\eqn{classbr}. The Poisson map $\pi$ is {\it not\/}
required to be onto, nor a submersion, that is to say, to
arise from a regular foliation of $\real^3$.

Several $\pi$-maps were constructed in~\cite{MMVZ}, under the name of
(generalized) classical Jordan--Schwinger maps. We give just a set of
possible realizations of the maps $f$, together with a Casimir function
$\pi^*{\cal C}$.

Let us consider first the cases A, in which the form
$\alpha=d{\cal C}$ is exact.

\begin{itemize}

\item[A.1]
$su(2)$
$$
f_1 \equiv \pi^*x = \half(q_{1}q_{2}+p_{1}p_{2}),~~~ f_2 \equiv
\pi^*y = \half(q_{1}p_{2}-q_{2}p_{1}),$$
\be f_3 \equiv \pi^*w =
\fr{1}{4}(q_{1}^2 + p_{1}^2 - q_{2}^2 - p_2^2)\label{jssu}
\ee
satisfying the relations
$$
\{\pi^*x,\pi^*y\} = \pi^*w,~~~~    \{\pi^*y,\pi^*w\} = \pi^*x,$$
\be
\{\pi^*w,\pi^*x\} = \pi^*y.    \label{su2}
\ee
The Casimir function $\frac{1}{2}(f_1^2 + f_2^2 + f_3^2)$ is given by
$\pi^*{\cal C}=\frac{1}{32}(p_{1}^2 + q_{1}^2 + p_{2}^2 + q_2^2)^2$.

\item[A.2] $e(2)$
$$
f_1\equiv\pi^*x = \half (q_2^2+p_2^2-q_1^2-p_1^2),~~~~
f_2\equiv\pi^*y = q_2+p_1,$$
\be
f_3\equiv\pi^*w = -q_1-p_2  \label{jse2}
\ee
with
$$
\{\pi^*x,\pi^*y\} = \pi^*w,~~~~    \{\pi^*y,\pi^*w\}=0,$$
\be
\{\pi^*w,\pi^*x\} = \pi^*y.    \label{e2}
\ee
The Casimir function $\frac{1}{2}(f_2^2 + f_3^2)$ is given
by $\frac{1}{2}(p_{1}^2 + q_{1}^2 + p_{2}^2 + q_2^2 +2p_1q_2
+ 2p_2q_1)$.

\item[A.3] $sl(2,\real)$
$$
\pi^*x = \fr{1}{4}(p_{1}^2 + q_{1}^2 + p_{2}^2 + q_2^2),~~~~
\pi^*y = \fr{1}{4}(q_{1}^2 + q_{2}^2 - p_{1}^2 - p_2^2),$$
\be
\pi^*w = \half(p_{1}q_{1} + p_{2}q_{2})   \label{jssl}
\ee
with
$$
\{\pi^*x,\pi^*y\} =-\pi^*w,~~~~    \{\pi^*y,\pi^*w\}=\pi^*x,$$
\be
\{\pi^*w,\pi^*x\}=-\pi^*y.    \label{sl2}
\ee
The Casimir function $\frac{1}{2}(f_1^2 - f_2^2 - f_3^2)$
yields $\frac{1}{8}(q_{1}p_{2}-q_{2}p_{1})^2$.

\item[A.4] $iso(1,1)$
\be
\pi^*x=\half(p_1^2+p_2^2-q_1^2-q_2^2),~~~~
\pi^*y=p_1+q_2,~~~~
\pi^*w=-p_2-q_1\label{jsiso}
\ee
with
$$
\{\pi^*x,\pi^*y\} =\pi^*w,~~~~    \{\pi^*y,\pi^*w\}=0,$$
\be
\{\pi^*w,\pi^*x\}=-\pi^*y.    \label{iso11}
\ee
and the Casimir function $\frac{1}{2}(- f_2^2 + f_3^2)$ is
$\frac{1}{2}(q_1^2-q_2^2-p_1^2+p_2^2+2q_1p_2-2q_2p_1)$.

\item[A.5] $h(1)$
\be
\pi^*x=q_1~~~~\pi^*y=p_{1}q_{2}~~~~\pi^*w=q_2     \label{jsh1}
\ee
satisfying
\be
\{\pi^*x,\pi^*y\} =\pi^*w,~~~~    \{\pi^*y,\pi^*w\}=0,~~~~
\{\pi^*w,\pi^*x\}=0.    \label{h1}
\ee
The Casimir function $\frac{1}{2}(f_3^2)$ is $\pi^*{\cal
C}=\frac{1}{2}q^2_{2}$.

\smallskip

The trivial case of the abelian algebra, obtained by taking all the
coefficients in \eqn{classbr} equal to zero, may be realized by many
functions.

\smallskip

We pointed out already that the maps $\pi$ are not
onto, in general. We have an onto map in
the $su(2)$ case, but, for instance, for the $sl(2,\real)$ case, $\real^4$ is
projected onto a solid cone.

\smallskip

\item[B.] The remaining cases, for which $\alpha$ is not closed,
can be treated in a uniform way by giving the realization
\be
\pi^*x=-h(q_1p_1+q_2p_2) - cq_2p_1 + bq_1p_2,~~~~\pi^*y=q_1,~~~~
\pi^*w=q_2 \label{jssb}
\ee
satisfying the relations
$$
\{\pi^*x,\pi^*y\} = h\pi^*y + c\pi^*w,\;   \{\pi^*y,\pi^*w\}=0,$$
\be
\{\pi^*w,\pi^*x\} = b\pi^*y - h\pi^*w.    \label{sb2}
\ee
In this case there is no Casimir function, as the form
$\alpha=h(q_1dq_2-q_2dq_1) +bq_1 dq_1 +c q_2 d q_2$ is not closed.
\end{itemize}

The reader will have noticed that all realizations chosen are by means of
quadratic and linear functions of the canonical coordinates; of course,
essentially equivalent ones may be obtained through linear canonical
transformations. However, it should be pointed out that there are not the
only possible ones: there exist many other inequivalent realizations by
different classes of functions. For instance, there is for case B.1 the
alternative realization:
\be
\pi^*x=-p_{1}-p_{2},~~~~\pi^*y=e^{q_{1}},~~~~
\pi^*w=e^{q_{2}}. \label{jssbbis}
\ee
This form can be obtained from the previous one with a (singular) canonical
transformation:
\begin{eqnarray*}
p^\prime_i&=&p_iq_i\\ q^\prime_i&=&\log q_i.
\end{eqnarray*}

\section{The main principle}

We steer now to prove that novel noncommutative products can be obtained on
$\real^3$ from the Moyal product in four dimensions, via the Poisson maps
described in the previous section. The Moyal product is described in the
appendix, where we also set notations and collect some other relevant
material. In the next section explicit formulae for the $\ast_{\cal G}$ are
exhibited. The formulae themselves attest in each case that the star
product in $\real^4$ of functions of the $\pi^*x,\pi^*y,\pi^*w$ variables
depends only on the $\pi^*x,\pi^*y,\pi^*w$ variables, and an easy induction
argument shows this to be the case in all instances. However, the more
mathematically minded readers may prefer an {\it a priori} argument. That
we give here.

Consider the noncompact space $\real^3$ with coordinates $x,y,w$. We want
to define a deformed product $\ast_\g$ of the algebra of functions of $x,y$
and $w$, with the property:
\be
[x,y]_{\ast_\g} = i\theta\,\{x,y\}_\g,
\ee
for $\g$ any of the Lie algebras obtained in Section~2.

Let us assume that a nonzero vector field $H$ exists such that
\be
L_H \pi^*x=L_H \pi^*y=L_H \pi^*w=0,
\label{predefI}
\ee
with $L_H$ the Lie derivative in the direction of $H$. Let $z$ be a
convenient (local, if you wish) fourth coordinate on $\real^4$. It is
impossible that $L_Hz=0$, too. This means that, given any function
$F(x,y,w)$, we can characterize $\pi^*F$ as a function on $\real^4$ by the
fact that
\be
L_H\pi^*F=0.
\label{predefII}
\ee
The crucial condition to meet then is that, given $F(x,y,w), G(x,y,w)$, it
obtain
\be
L_H(\pi^*F\star_\theta\pi^*G) = 0,
\label{stardef}
\ee
where $\star_\theta$ denotes the Moyal product.%
\footnote{Often, when this is not too
liable to confusion, the suffix $\theta$ is omitted from the notation,
likewise we will often suppress the ${\cal G}$ from the $\ast_{\cal G}$.}

In effect, fulfilling equation~\eqn{stardef} ensures that the following
procedure is well defined: to multiply two functions of $x,y,w$, lift them
to four dimensions to obtain functions on $\real^4$ that can be multiplied
with the four dimensional Moyal $\star_\theta$ product. The resulting
function will still be in the kernel of $H$, and it is therefore possible
to regard it as a function on the three dimensional space. In other words:
when we $\star$-multiply two functions belonging to the Funktionengruppe,
the result still belongs to the Funktionengruppe, and:
\be
\pi^*(F\ast_\g G)=\pi^*F\star\pi^*G. \label{prodg}
\ee
defines $F\ast_\g G$.

Vector fields with property~\eqn{predefI} do exist. Consider the vector
field $X$ defined by
\be
i_X\omega = -\pi^*\alpha.
\ee
In cases A, this is just the Hamiltonian vector field associated to the Casimir
function $\pi^*\cal C$. (We leave aside the trivial case in which $X=0$.) By
construction~\eqn{predefI} holds with $H=X$.

To prove~\eqn{stardef}, we consider several situations.

We dispose first of the case of the abelian Lie algebra. Then the star
product becomes the ordinary product and then of course the
Funktionengruppe closes.

For the cases A with quadratic Casimir, since
$L_X(\pi^*F)=\{\pi^*F,\pi^*{\cal C}\}$, and in view of
equation~\eqn{eq:Moyal-Poissonmagic} of the appendix,
property~\eqn{stardef} can be rephrased as
$$
[\pi^*F,\pi^*{\cal C}]_\star = [\pi^*G,\pi^*{\cal C}]_\star = 0\quad{\rm
implies}\quad [\pi^*F\star\pi^*G,\pi^*{\cal C}]_\star = 0,
$$
which is obvious.

For the cases A with quartic Casimir, we can consider instead the
Hamiltonian vector field $H$ associated to the square root of $\pi^*\cal
C$, which has the same properties. The result follows from the same
argument. We remark that the idea is already present in~\cite{HLS}.

In some cases, that line of reasoning in terms of the Lie derivative with
respect to the vector field $H$ can be recast in pure Poisson algebra
terms.

For instance, for the $su(2)$ case, we assert: the Poisson subalgebra
generated by the quadratic functions $\pi^*x,\pi^*y,\pi^*w,f_H$ (with $f_H$
the Hamiltonian function associated to H) is the {\it Poisson commutant\/}
of $f_H=q_1^2+q_2^2+p_1^2+p_2^2$. In effect, by virtue of the Lie algebra
isomorphism discussed in the appendix, the problem reduces to the simple
linear algebra exercise of finding the centralizer of the matrix $J$
of~\eqn{laJ} in $sp(4;\real)$. This centralizer is given by
\[
B = \left(\begin{array}{cccc}
0 & a & b & c \\ -a & 0 & c & d \\ -b & -c & 0 & a \\ -c & -d & -a & 0
\end{array}\right).
\]
By the way, it is isomorphic to $u(2)$. The quadratic form ${}^tuBu$ then
reproduces the functions $\pi^*x,\pi^*y,\pi^*w,f_H$. The claim is proved.

But such a commutant is bound to be an (involutive) {\it Moyal} subalgebra,
as $\{f_H,F\} = \{f_H,G\} = 0$ imply $\{f_H, F\star_\theta G\}=0$, in view
of~\eqn{eq:Moyal-Poissonmagic} again. As $q_1^2+q_2^2+p_1^2+p_2^2$ is a
function of $\pi^*x,\pi^*y,\pi^*w$ itself, we are done. The latter kind of
argument goes back to the work by Bayen {\it et al\/}~\cite{Bayenetal}, who
reintroduced the Moyal product in theoretical physics.

For the cases B, the proof is slightly more involved. To make it clearer,
let us look at the simpler case B.1.

{}First we reconsider the previous reasoning. They are but an infinitesimal
version of the fundamental fact that linear symplectic transformations are
implemented by inner automorphisms of the Moyal algebra. The implementers
of those automorphisms are exhibited in the appendix in a fully explicit
way. Now, just observe that $\pi^*x,\pi^*y,\pi^*w$ in case $B.1$ generate
the commutant of $\arctan(q_1/q_2)$. While the canonical transformations
generated by the latter function are not linear, they still are implemented
by inner automorphisms of the Moyal algebra. This happens because they give
rise to bicanonical maps in the sense of Amiet and Huguenin~\cite{AmiHu}
for the Moyal product. The detailed treatment of the bigger class of
implementors would take us too far afield, however, and is left for another
paper.

The validity of the formulae in the next section is not restricted by the
fact that the image of the ``scaffolding map" $\pi$ can eventually be less
than all of $\real^3$. Associativity of the $\ast_{\cal G}$ and the
involution are inherited from the Moyal product.

The star products can be seen to give realizations of the
enveloping Hopf algebras $U(su(2)), U(sl(2,\real))\dots$  At
the abstract level, it has been clear for a long time that
the ``quantization" of a linear Poisson structure is given
by the universal enveloping algebra of the corresponding Lie
algebra~\cite{GinebraplusJano}. No big deal, then, except that
we now possess a concrete, not just formal, realization-cum-completion of
those algebras, which hopefully will allow a fresh attack on
the long-standing problem of the existence of (metric)
noncommutative geometries over them.

\section{The new star products}

We now illustrate the new deformed products obtained from
the Moyal product on $\real^4$ through the reduction maps
$\pi$ defined in Section~{\bf 3}. Each algebra of functions
on $\real^3$ identified by the reduction map is closed with
respect to the relative star product. When relevant, we also
discuss representations and other notable aspects of the
resulting algebras. We have no pretense in this section to
give an exhaustive treatment of these products, and limit
ourselves to explicit expressions of the products, as well
as some other aspects considered relevant.

As an initial remark, note the similarity, advertised in the
Introduction, of our procedure with the transition from the
Moyal product to the noncommutative tori
$\torus^{n}_\theta$; the latter is identified also as a
subalgebra of the Moyal algebra, associated with the
invariance under an action of the group $\zed^n$. We outline
how this can rigorously be proved, and refer to~\cite{Metis}
for a more detailed treatment.

The Gelfand $C^*$-algebra representing the real plane is the algebra
$C_0(\real^2)$ of continuous functions on $\real^2$, vanishing at infinity.
This can be compactified (unitized) in several ways. The {\it maximal\/}
compactification is given by the algebra $C_b(\real^2)$ of continuous and
bounded functions on $\real^2$; this coincides with the {\it multiplier\/}
algebra~\cite[Sect.~1.3]{Polaris}: $fg\in C_0(\real^2)$ for all $g\in
C_0(\real^2)$ iff $f\in C_b(\real^2)$. For all practical purposes one needs
to work in the smooth category, so one is led to consider the dense
Schwartz $\cal S$ subalgebra of smooth functions rapidly vanishing at
infinity, together with all their derivatives, and the dense subalgebra of
$C_b$ of bounded smooth functions, all of whose derivatives are bounded.
The latter space is denoted ${\cal O}_0$ in distribution theory. It is a
subspace of the space ${\cal O}_{c}$ of polynomially bounded smooth
functions, together with all their derivatives, with the degree of the
polynomial bound independent of the derivative.

Endow $\real^2$ with the Moyal product~$\star$. It turns out~\cite{Phobos}
that ${\cal S}\star_\theta{\cal S} ={\cal S}$ for all $\theta$ and (with an
appropriately extended definition of $\star$) that ${\cal O}_{0}\star{\cal
S} = {\cal S}\star{\cal O}_{0} = {\cal S}$. Intuitively evident, and true,
but a bit harder to prove, is that ${\cal O}_{c}\star{\cal
O}_{c}\subset{\cal O}_{c}$ and ${\cal O}_{0}\star{\cal O}_{0}\subset{\cal
O}_{0}$. A proof was given long ago by Figueroa%
\footnote{A Fourier-transformed version of Figueroa's result has appeared
recently in~\cite{MormonChurch}.}
~\cite{Amalthea}. Therefore ${\cal O}_{0}$ is not only a
compactification of the {\it commutative} plane, but also a
compactification of the {\it Moyal} plane.

The Moyal star is invariant by translation. Therefore the product of two
periodic functions with a fixed period is also periodic with the same
period. The subalgebras of periodic elements in ${\cal O}_{0}$ coincide
with the smooth noncommutative torus algebras. End of the remark.

As we argued in the previous section, the definition of the
new products is given by
\be
\pi^*(F\ast_\g G)=\pi^*F\star\pi^*G.
\ee
The explicit formulae for the new products are easily computed by means of
the expansion of the Moyal product (equation~\eqn{eq:pract-prodform} in the
appendix). We introduce the notation $f(x_i*)$ for functions on $\real^3$,
meaning they have to be intended as expansions in $\ast$-powers of the
arguments. An analogous notation is used for functions on $\real^4$,
meaning that they have to be intended as an expansion in $\star$-powers of
the arguments. The Moyal product defined in~\eqn{eq:pract-prodform} can be
then rewritten in the language of pseudodifferential operators under the
form
\be
f(q^j\star, p^j\star) \star g(q^j, p^j)= f(q^j + {i\theta\over
2}{\del\over\del p^j}, p^j -{i\theta\over 2} {\del\over\del
q^j})g(q^j,p^j).
\label{expostar}
\ee

\subsection{$su(2)$}
The Poisson subalgebra generated by the quadratic functions $\pi^*x_i$,
in~\eqn{jssu} is the commutant of the space of functions of $f_H$ with
$f_H= p_1^2+q_1^2+p_2^2+ q_2^2$, essentially the square root of the
Casimir. By means of~\eqn{eq:pract-prodform} and~\eqn{expostar} we find:
\be
x_j \ast_{su(2)} f(x_i) = \{x_j -{i\theta\over 2}\epsilon_{jlm} x_l \del_m
- {\theta^2\over 8}[(1 + x_k \del_k) \partial_j
- {1\over 2} x_j \,\del_k \del_k]\} f(x_i).
\label{starsu2}
\ee
{}From this, general formulae similar to~\eqn{expostar} are
obtained. And so, the Casimir function in the sense of
enveloping algebras is
\be
{\cal C}(x_i*) = \half(x\ast x+ y\ast y +w\ast w) =
\half(x^2+y^2+w^2-\fr{3}{8}\theta^2).
\ee

The function $f_H$ has been recognized as the Hamiltonian of the
two--dimensional harmonic oscillator. It is therefore natural to define the
complex quantities
\be
z_i=q_i+ip_i
\ee
with the Poisson brackets $\{z_i,\bar z_j\}=-2i\delta_{ij}$.
In this case the four dimensional $\star$ product becomes
the operator product of the functions of the well known
$a_i,a^\dagger_i$ of the harmonic oscillator, with $\theta$
playing the role of $\hbar$; this realizes in fact the
Jordan-Schwinger map. A presentation of the Lie algebra
$su(2)$ in terms of creation and annihilation operators,
$a_i, a^{\dag}_i,~i\in\{1,2\}$, is provided by
\be
X_{1} = \frac{i(X_{+} + X_{-})}{2},~~~~~~~~X_{2} = \frac{X_{-} -
X_{+}}{2},~~~~~~~~X_{3}= iX_{0},
\ee
with
\be
X_{+} = a_{1}^{\dag}a_{2}~,~~~~~~~~ X_{-} =
a_{2}^{\dag}a_{1}~,~~~~~~~~X_{0} =
\fr{1}{2}(a_{1}^{\dag}a_{1} - a_{2}^{\dag}a_{2}).
\ee
We can have these functions act on the usual Hilbert space of the two
dimensional harmonic oscillator, with basis the cartesian kets:
$\ket{n_1,n_2}$. The generators act as:
\beq
X_+\ket{n_1,n_2}&=&\theta\sqrt{(n_1+1)n_2}\ket{n_1+1,n_2-1}\nn\\
X_-\ket{n_1,n_2}&=&\theta\sqrt{(n_2+1)n_1}\ket{n_1-1,n_2+1}\nn\\
X_0\ket{n_1,n_2}&=&\theta(n_1-n_2)\ket{n_1,n_2};
\label{Xact}
\eeq
notice that the ``energy`` $n_1+n_2$ does not change (a
consequence of the form of the Casimir), so that is it
natural to change the basis and choose a basis of
eigenstates of the Hamiltonian and the angular momentum.
Define
\be
\psi_{lm}\equiv\ket{l+m,l-m}~~~~~~~~~~~~
\ket{n_1,n_2}\equiv\psi_{\frac{n_1+n_2}2,\frac{n_1-n_2}2}
\ee
with $l\geq 0,\ -l\leq m \leq l$. Then \eqn{Xact} becomes
\beq
X_{\pm}\,\psi_{l,m}&=&\theta\sqrt{l(l+1)-m(m{\pm}1)}\,\psi_{l,m{\pm}1}\nn\\
X_0\,\psi_{l,m}&=&\theta l(l+1)\,\psi_{l,m}.
\eeq
For each value of $l$ (integer or half integer) there is a representation
of $su(2)$. The algebra of functions of $\real^3$ therefore reduces to a
set of finite dimensional algebras, receptacles for representations of
$su(2)$. This algebra can be given an interesting interpretation. Each
reduced block is the algebra of a fuzzy sphere~\cite{fuzzy1,Miranda}, of
radius $\theta\sqrt{l(l+1)}$ in the oscillator representation. Therefore
the three dimensional space is ``foliated'' as a set of fuzzy spheres of
increasing radius.

It is truly remarkable that, conversely, the {\it Moyal\/} product can be
obtained from the fuzzy sphere of~\cite{Miranda}, by means of group
contraction from $SU(2)$ to the Heisenberg group~\cite{AmietW}.

We can give a geometric interpretation of the new star product. Note that,
with the exception of the zero orbit, the orbits of the Hamiltonian system
associated to $f_H$ are circles. Functions of $(x,y,w)$ correspond here to
functions of $(q_1,q_2,p_1,p_2)$ that remain invariant on those orbits. We
are thus identifying $\real^3$ to the foliation of $\real^4$ by those
trajectories. The orbits are all parallel and rest on spheres in $\real^4$.
One circle and only one passes through each point different from 0. The
corresponding maps $S^3\to S^2$ are Hopf fibrations.

Recall that a three dimensional product related to the fuzzy sphere has
been introduced in~\cite{HLS}. Although similar, the product described in
this subsection is not identical to the one introduced by Hammou and
coworkers. For instance, one can check that, if $r=\sqrt{x^2+y^2+w^2}$,
then $x_j\ast_{su(2)} r = r\ast_{su(2)} x_j$ equals simply $rx_j$, which is
not the case for their product. The difference is attributable to the
product in~\cite{HLS} being based on normal ordering, while we use Weyl
ordering. Except for that, it can be regarded as a particular case of our
construction.

\subsection{$e(2)$}
In this case the Casimir function is quadratic, hence the
Poisson subalgebra generated by the quadratic functions
$\pi^*x_i$ \eqn{jse2} is the Poisson commutant of
$\pi^*{\cal C}$. The correspondent star product is found to
be
\beq
x\ast_{e(2)}f(x,y,w) &=&
\{x-{i\theta\over2}(y\partial_w-w\partial_y) -
\quad {\theta^2\over 8}[4\partial_x +2x \partial_x^2 +
w\partial_x\partial_w+ y\partial_x\partial_y ]\} f
\nn\\
y\ast_{e(2)}f(x,y,w) &=&
\{y-{i\theta\over2}w\partial_x \} f
\nn\\
w\ast_{e(2)}f(x,y,w) &=&
\{w+{i\theta\over2}y\partial_x \} f.
\eeq
The Casimir function is
\be
{\cal C}(x_i*) = \half(y\ast y +w\ast w) = \half(y^2+w^2),
\ee
and it coincides with the ordinary Casimir. The representation theory can
be obtained from that of $su(2)$ through standard contraction techniques.

\subsection{$sl(2,\real)$}
As for the $su(2)$ case, the Casimir is here quartic. Then  the Poisson
subalgebra generated by the functions $\pi^*x_i$  is the commutant of
$f_H=q_1p_2-q_2p_1$, the square root of the Casimir function. The star
product induced by $sl(2,\real)$ through the maps~\eqn{sl2} is:
\be
x_j \ast_{sl(2,\real)} f(x_i) = \{x_j - {i\theta\over 2}g_{jk}c_{klm} x_l
\del_m  - {\theta^2\over 8}[(1 + x_l \del_l) g_{jk}\partial_k
- {x_j\over 2}g_{lm}\del_l\del_m]\} f
\label{starsl2R}
\ee
where $g_{ij}=diag(1,-1,-1)$ is the invariant metric in $sl(2,\real)$. It
is formally identical to the one relative to $su(2)$, substituting $g_{ij}$
for $\delta_{ij}$. The Casimir function is:
\be
{\cal C}(x_i*)={1\over 2}(x\ast x- y\ast y -w\ast w)={1\over
2}(x^2-y^2-w^2-{3\over 8 }\theta^2).
\ee
The orbits of the angular momentum $f_H$ are also circles, but the
quotient map from $\real^4$ to the 3-manifold with boundary $\{(x,y,w):
x^2-y^2-w^2 = {f_H}^2\geq0\}$ is topologically trivial.

\subsection{$iso(1,1)$}
This case is similar to the Euclidean case. The Casimir is quadratic and
the star product induced by the Poincar\'{e} algebra is
\beq
x\ast_{iso(1,1)}f(x,y,w) &=&
\{x+{i\theta\over2}(y\partial_w+w\partial_y) +
\quad {\theta^2\over 8}[4\partial_x +2x \partial_x^2 +
w\partial_x\partial_w+ y\partial_x\partial_y ]\} f
\nn\\
y\ast_{iso(1,1)}f(x,y,w) &=&
\{y-{i\theta\over2}w\partial_x \} f
\nn\\
w\ast_{iso(1,1)}f(x,y,w) &=&
\{w-{i\theta\over2}y\partial_x \} f.
\eeq
The Casimir function is:
\be
{\cal C}(x_i*) = \half(- y\ast y +w\ast w) = \half(-y^2+w^2).
\ee
It is local and coincides with the ordinary Casimir.

\subsection{$h(1)$}

We now discuss the case associated to the Heisenberg Lie algebra, whose map
from $\real^4\to\real^3$ is~\eqn{jsh1}. In this map the quantity $p_2$
never appears, and therefore the algebra generated by these generators is
highly reducible (not surprisingly given the essential unicity of the
representations of the Heisenberg algebra), and evaluated at a particular
$w$ it reduces simply to a copy of the Moyal algebra. The star product is
given by:
\beq
x\ast_{h(1)}f(x,y,w) &=&
\{x+{i\theta\over2}w\partial_y \}f \nn\\
y\ast_{h(1)}f(x,y,w) &=&
\{y-{i\theta\over2}w\partial_x \} f \nn\\
w\ast_{h(1)}f(x,y,w) &=& w f.\nn
\eeq

The Casimir function
\be
{\cal C}(x_i*)=\half(w\ast w)=\half w^2
\ee
coincides with the ordinary Casimir. In~\cite{LSZ}, an explicit expression
for the product of two arbitrary monomials is given.

The three dimensional space in this case is given by the
space spanned by $q_1,q_2$ and $p_1$~, with the choice
of~\eqn{jsh1}. This is in turn foliated by the product of a
line ($q_2$) and a set of Moyal planes spanned by $p_1q_2$
and $q_1$. Note that the plane $w=0$ has different
properties, as it is well known from the theory of
representations of the Heisenberg algebra.

\subsection{{\rm B: the algebras} ${\cal G}_h$}

These algebras corresponding to the nonclosed Casimir form may be treated
in a unified manner. The star product induced by the quadratic realization
of ${\cal G}_h$ \eqn{sb2} is:
\beq
x\ast_{{\cal G}_h} f(x,y,w) &=&
\{x+{i\theta\over 2}[ h (y\partial_y+w\partial_w)+cw\del y -b y\del w] - \nn\\
&&\quad {\theta^2\over 4}(h^2+bc) [2\partial_x +x \partial_x^2
+ w\partial_x\partial_w+
y\partial_x\partial_y ]\} f\nn
\\
y\ast_{{\cal G}_h} f(x,y,w) &=&
\{y-{i\theta\over 2} (h y+c w)\partial_x \} f \nn\\
w\ast_{{\cal G}_h} f(x,y,w) &=&
\{w-{i\theta\over 2}(h w- by) \partial_x \} f.
\eeq
This star product is not equivalent to the one obtained from the non
linear realization \eqn{jssbbis}. Concentrate on the $sb(2,\complex)$
case. Here $h=1$, $b=c=0$. The representation of this algebra on
$L^2(\real^2)$ is reducible, but it is indecomposable, a trademark of
triangular algebras. The Weyl map on $L^2(\real^2)$ of the
generators~\eqn{jssb} yields:
\be
\hat X = -\frac{1}{2}\sum_{i=1,2}(\hat p_i\hat q_i+\hat q_i\hat p_i),~~~~~
\hat Y = \hat q_1,~~~~~ \hat W=\hat q_2,
\ee
where the $\hat{~}$ denotes usual quantum operators. Consider the
(nonorthonormal) basis:
\be
\psi_{nm}=q_1^nq_2^me^{-(q_1^2+q^2_2)}
\ee
and define ${\cal H}_{NM}$, the Hilbert subspaces spanned by $\psi_{nm}$
with $n>N,m>M$. It is easy to see that these spaces are left invariant by
any function of the operators $\hat X,\hat Y$ and $\hat W$ for any choice
of $N$ and $M$. The representation is therefore reducible. It is not
however possible to block--diagonalize it, and it is therefore
indecomposable.

\smallskip

The presentation of the precedent star algebras by unitaries
and relations is a straightforward (if somewhat tedious)
task. See the discussion in the appendix; for further
detail, the reader is advised to consult~\cite{TritonAriel}.

\smallskip

It is interesting to compare our products with the formal products
introduced by Kontsevich~\cite{Maxim}. Both are $\cal G$-covariant. One can
see that higher order cochains contain exactly the same type of terms, but
the weights are different. Very likely, the products are equivalent in the
very general sense of~\cite{Maxim}. Our products differ in principle from
Kontsevich's in that they are non-formal; equivalence in the formal sense
does not signify representations on Hilbert space (if any) are equivalent.
Also, the operators involved in the asymptotic expansion in $\theta$ for
our products are not necessarily (bi-)differential, they can be
pseudodifferential.

It would also be interesting to know how the present star
products are related to the foliation algebras introduced by
Connes~\cite{CplusDND}.

\section{Noncompact star triples}

In this section and the next we lay the foundations of a theory of
noncompact spectral triples. Although the problems are mathematical, our
approach will be guided by a more utilitarian view proper to physicists.

A {\it compact} spectral triple is a triple $(\A,\hil,D)$, where $\A$ is a
unital pre-$C^*$-algebra, $\hil$ is a Hilbert space carrying a
representation of~$\A$ by bounded operators, and $D$ is a selfadjoint
operator on~$\A$, with compact resolvent $R_D(\lambda):=(D-\lambda)^{-1}$,
such that the commutator $[D,a]$ is also bounded on~$\hil$, for each $a \in
\A$.

Spectral triples come in two parities, odd and even. In the odd case, there
is nothing new; in the even case, there is a grading operator $\chi$ on
$\hil$ (a bounded selfadjoint operator satisfying $\chi^2 = 1$), such that
the representation of $\A$ is even and the operator $D$ is odd; thus each
$[D,a]$ is a bounded odd operator on~$\hil$.

In the noncompact case we need to modify this definition, as $\A$ will no
longer be unital, and, more to the point, the commutative example indicates
that compactness of $R_D(\lambda)$ is out of question. The generalization
was pointed out by Connes himself in~\cite{realncg}: we ask $aR_D(\lambda)$
to be compact for any $a\in\A$.

As hinted in the Introduction, noncommutative compact spin geometries are
spectral triples satisfying several extra conditions, arising from
consideration of the algebraic properties of ordinary spin and metric
geometry. Seven such properties were put forward in~\cite{ConnesGravplus}.
Here we just summarize them; a more complete account is given
in~\cite[Sect.~10.5]{Polaris}.

\begin{enumerate}

\item{\it Classical dimension\/}:
There is a unique nonnegative integer $n$, the ``classical dimension'' of
the geometry, for which the eigenvalue sums $\sg_N := \sum_{0\leq k<N}
\mu_k$ of the compact positive operator $|D|^{-n}$ satisfy $\sg_N \sim C
\log N$ as $N \to \infty$, with $0 < C < \infty$. The coefficient is written
$C = \int |D|^{-n}$, where $\int$ denotes the Dixmier trace
if $n \geq 1$; it coincides essentially with the Wodzicki
residue~\cite[Chapter 7]{Polaris}. This $n$ is even if and
only if the spectral triple is even. (When $\A =
C^\infty(M)$ and $D$ is a Dirac operator, $n$ equals the
ordinary dimension of the spin manifold~$M$).

\item{\it Regularity\/}:
Not only are the operators $a$ and $[D,a]$ bounded, but they lie in the
smooth domain of the derivation $\delta(T) := [|D|,T]$. (When $\A$ is an
algebra of functions and $D$ is a Dirac operator, this smooth domain
consists exactly of the $C^\infty$ functions.)

\item{\it Reality\/}:
There is an antiunitary operator $C$ on~$\hil$, such that $[a, Cb^*C^{-1}]
= 0$ for all $a,b \in \A$ (thus $b \mapsto Cb^*C^{-1}$ is a commuting
representation on $\hil$ of the ``opposite algebra'' $\A^\circ$, with the
product reversed). Moreover, $C^2 = {\pm} 1$, $CD = {\pm} DC$, and $C\chi
= {\pm}\chi C$ in the even case, where the signs depend only on $n
\bmod 8$. (In the commutative case, $C$ is the charge conjugation operator
on spinors.)

\item{\it First order\/}:
The bounded operators $[D,a]$ commute with the opposite algebra representation:
$[[D,a], Cb^*C^{-1}] = 0$ for all $a,b \in \A$.

\item{\it Finiteness\/}:
The algebra $\A$ is a pre-$C^*$-algebra, and the space of smooth vectors
$\hil_\infty := \bigcap_k Dom(D^k)$ is a finitely generated projective left
$\A$-module. (In the commutative case, this yields the smooth spinors.)

\item{\it Orientation\/}:
There is a \textit{Hochschild $n$-cycle} $c$, on~$\A$ with values in
$\A\otimes\A^\circ$, whose natural representative is denoted $\pi_D(c)$.
Such an $n$-cycle is usually a finite sum of terms like $(a\otimes b)
\otimes a_1 \otimes\cdots\otimes a_n$ which map to operators
\be
\pi_D((a\otimes b) \otimes a_1 \otimes \cdots \otimes a_n) :=
aCb^*C^{-1} \,[D,a_1] \dots [D,a_n],
\ee
and $c$ is the algebraic expression of the \textit{volume form} for the
metric determined by~$D$. This volume form must solve the equation
\be
\pi_D(c) = \chi \mbox{ \ (even case), or}\quad
\pi_D(c) = 1    \mbox{ \ (odd case)}.
\label{Hochschildeq}
\ee

\item{\it Poincar\'{e} duality\/}:
The index map of~$D$ determines a nondegenerate pairing on the
$K$-theory of the algebra~$\A$. This is related to the existence of
Morita duality between $\A$ and the ``quantum Clifford algebra"
constructed from $\A$ and $D$~\cite{RennieII}. (In the commutative case,
the Chern homomorphism matches this nondegeneracy with Poincar\'{e}
duality in de~Rham (co)homology.)

\end{enumerate}

For proofs of the fact that when $A = C^\infty(M)$ the usual apparatus of
geometry on spin manifolds (spin structure, metric, Dirac operator) can be
fully recovered from these seven conditions in the compact case,
see~\cite{RennieI} and~\cite[Chap.~11]{Polaris}.

The question is now to modify these conditions for our
needs, since we deal with noncompact manifolds. This we do
guided more by physical insight than by pretensions of
rigour; all examples we consider have the topology of
Euclidean spaces. In the Moyal case, whenever we need to fix
ideas, and in consonance with the first part of the paper,
we think of~${\real}_\theta^4$.

The minimum requirement for new postulates, beyond being consistent with
the axioms for compact spaces, is to cover the commutative case. In
particular, we give ourselves just the task of checking whether the linear
spaces $\real^n$ with their usual spin structure satisfy the conditions
laid down below. The inverse task of recovering that spin structure from
the axiom is to follow the pattern established in~\cite{RennieI}
and~\cite{Polaris}. It is left for a better occasion. Related ideas have
been discussed in~\cite{RennieII}. We have already indicated that, when
needed, we deal with {\it star triples\/}, defined as deformations of
commutative spectral triples. The latter will be seen to cover as well the
Moyal product cases. This keeps great heuristic value: as a consequence,
the Chamseddine--Connes spectral action principle~\cite{ChC} can be
extended to the Moyal framework~\cite{Metis}.

For convenience, we restate the definition of noncompact
spectral triple, and carry the discussion of the ``strict"
and the ``star triple" framework in parallel.

We define: an odd {\it noncompact} spectral triple is a
triple $(\A,\hil,D)$, where $\A$ is a non-unital
pre-$C^*$-algebra, $\hil$ is a Hilbert space carrying a
representation of~$\A$ by bounded operators, and $D$ is a
selfadjoint operator on~$\A$, such that $fR_D(\lambda)$ is
compact and the commutator $[D,f]$ is also bounded
on~$\hil$, for each $f\in\A$.

An odd {\it noncompact star triple\/} is a triple
$(\A,\hil,D)$, where $\A$ is a non-unital pre-$C^*$-algebra,
given by a star product on some space of functions and/or
distributions on a spin${}^c$ manifold (in other words, we
consider also a commutative product on the vector space
$\A$), $\hil$ is a Hilbert space carrying a representation
of~$\A$ by bounded operators, and $D$ is a selfadjoint
pseudodifferential operator on the manifold, such that
$fR_D(\lambda)$ is compact for each $f\in\A$, and all
commutators $[D,f]$ are bounded on~$\hil$.

In the even case, a grading operator $\chi$ on $\hil$ is
added in the same way as before.

Thus the first hurdle is to prove compactness of the
operators $fR_D(\lambda)$ ---where the action of $f$ by left
star multiplication is understood. This typically is easy
for star product algebras: they often give rise to compact
---indeed trace class--- operators.

Both in the commutative and the Moyal case, take for $\A$ the algebra $\cal
S$ of Schwartz functions on ${\real}^{2n}$; for $\hil$, a direct sum of
copies of the space of square summable functions on ${\real}^{2n}$. More to
the point, $\hil$ is the space of spinors on ${\real}^{2n}$, topologically
trivial. Finiteness and reality postulates prevent the Hilbert space on
which $\A$ is represented from being ``too small". We naturally want our
spectral triples be irreducible in some sense; but it is important to
realize that the $\hil$ must carry, as well as the representation of $\A$,
the one of $\A^\circ$ and a sundry list of operators.

The action of $\A$ by the Moyal star product on $\hil$ is
diagonal, so for analytical purposes we can assume only one
copy of $L^2({\real^{2n}})$ is present. The operators $Lf:
g\mapsto f\star g$ are trace class, with $$\Tr Lf =
(2\pi\theta)^{-n}\int_{{\real^{2n}}}f.$$ Perhaps the
simplest way to see that is to remember that the Wigner
transform intertwines $Lf$ with a multiple of the
Schr\"{o}dinger-like representation of $\A$ on
$L^2(\real)$~\cite{Deimos}. To be precise, there exists a
linear map
$$
W:{\cal S}({\real}^{2n}) \to
{\cal S}({\real}^{n})\otimes{\cal S}({\real}^{n})
$$
(unitary for an appropriate inner product) such that
$$
W(Lf)W^{-1} = \pi_S(f)\otimes I,
$$
where the action $\pi_S(f)$ is given by an integral kernel $k_f$ related to
$f$:
$$
[\pi_S(f)\varphi](x) = \int_{\real^n}k_f(x,y)\varphi(y)\,dy.
$$
In other words, $\pi_S(f)$ is the Weyl pseudodifferential operator
associated to the ``symbol" $f$. Moreover, $ W(f\star g) = Wf\circ Wg$,
where $\circ$ means composition of integral kernels. From the explicit form
of $\pi_S$ (see~\cite[sect.~3.5]{Polaris}):
\be
k_f(x,y) = \frac{1}{(2\pi\theta)^n}\int_{\real^n}
f\bigl(\frac{x+y}{2},z\bigr)\e^{i(x-y)z/\theta}\,dz
\label{WignerTr}
\ee
it is clear that
$$
\Tr Lf = \Tr\pi_S(f) =
\frac{1}{(2\pi\theta)^n}\int_{{\real}^{2n}} f(q,p)\,dq\,dp < \infty.
$$
Therefore, if $f\in{\cal S}$, then $fR_D(\lambda)$ is certainly compact.%
\footnote{By the way, the previous does {\it not\/} imply that $\pi_S(C_0)$ is the ideal of
compact operators.}

In the commutative limit, things are tougher, as neither $f$
nor $R_D(\lambda)$ are compact. However, the theorems quoted
in~\cite[Ch.~4]{BarryS} save the day. These refer to
operators (that appear naturally in scattering theory) of
the form $f(u)g(-i\nabla)$, with $\nabla$ the derivative
operator corresponding to the coordinates $u$, and count
among the more celebrated estimates in analysis.

Denote by $F$ the Fourier transform. To be precise, the
claim is that there exists a compact operator C such that,
for any two spinors $\Psi_1,\Psi_2$, with $F\Psi_2$ in the
domain of the multiplication operator $g$
$$
\<\Psi_1,C\Psi_2> = \<f\Psi_1,F^{-1}(gF\Psi_2)>.
$$
Now, for $g(D)=(D-\lambda)^{-1}$ on $\real^n$ we have $g\in
L^{n+1}({\real}^n)$. Also $f\in L^{n+1}({\real}^n)$,
obviously. Then $f(u)g(D)$ belongs to the Schatten class
$I^{n+1}$. This is clear for $n=1$, as we obtain a
Hilbert--Schmidt operator; and for $n=\infty$. The general
case follows by interpolation. Hence $f(u)g(D)$ is compact.

Now for the new postulates.

\begin{enumerate}

\item{\it Classical dimension\/}:
On $\real^n$, and, more generally, on a geodesically complete
spin manifold $M$ with Riemannian measure $\nu_g$, the operator $f|D|^{-n}$
belongs to the Dixmier trace class for all $f\in\A$, the trace being nonzero
in general. As the link with the Wodzicki residue {\it density}
is kept (see below), we can compute integrals by means of the
Dixmier trace $\Tr^+$:
$$
\int f(x)|\nu_g| = \frac{n(2\pi)^n}{\Omega_n}\Tr^+(f|D|^{-n}),
$$
where $\Omega_n$ denotes the volume of the sphere $\mathbb{S}^{n-1}$. 
Geodesical completeness is needed to make sure $D$ is selfadjoint. 
This formula appears already in~\cite[Corollary~7.22]{Polaris}.

The criterion fails spectacularly for the Moyal algebra
case: because $f$ is then trace class, its product with any
power $|D|^{-k}$ has vanishing Dixmier trace! This can be
understood as of the algebra $\cal S$ with the Moyal product
having vanishing dimension somehow; such a precipitous ``dimension drop"
indicates a severe difficulty for the transfer of geometrical features
from the commutative context, in the Moyal case. One could argue that
$\cal S$ is too ``small", when dealing with such product,
and define $\A$ as the ideal of elements $f$ such that $Lf$
of the Dixmier class. For $n=2$, the Moyal inverse of the
harmonic oscillator Hamiltonian (a gentle smooth function
vanishing not too rapidly at infinity) provides an example.
But such a procedure contains no information on the
dimension!

A replacement is at hand in the context of star triples, in
terms of the Wodzicki residue density. Let us just look at
the Dirac operator, and embrace the ``metric dimension"
definition. As $D$ is of commutative type, the only real
trouble is that, as the manifold is not compact, the
spectrum of $D$ is continuous. Then we must use a criterion
that makes no distinction between the discrete and the
continuous spectrum cases. Such a criterion was proposed
in~\cite{Odysseus} for (positive) elliptic
pseudodifferential operators. Let $A$ be one of such. The
idea is to look at the {\it spectral density\/} of $A$,
formally written as $\delta(\lambda-A)$. This is an
operator-valued distribution in ${\cal K}'$ (the space of
distributions that possess momenta at all orders: see the
definition of $\cal K$ a bit further on) with the property
$$
A^n = \int \lambda^n\delta(\lambda-A).
$$
Let $d_A(x,y;\lambda)$ denote the distributional kernel of
$\delta(\lambda-A)$. For the coincidence limit of the kernel $d_{|D|}$
(with $|D|$ of order 1) as $\lambda\uparrow\infty$, in dimension $n$ one
has:
\be
d(x,x;\lambda) \sim
\frac{1}{(2\pi)^n}\wres|D|^{-n}(x)\lambda^{n-1} + \cdots,
\label{Cesaro1}
\ee
where ``$\wres$'' denotes the Wodzicki residue density.%
\footnote{In case the reader gets nervous about the $|D|^{-n}$
notation, for the spectrum of $D$ may well include zero, it
can be replaced by $(D^2+\eps)^{-n/2}$ in all cases. In
other words, for simplicity we have assumed in the notation
that $D$ has a ``mass gap" around zero; but this assumption
we can easily dispense with.}
This we take as our replacement criterion: there must exist
a unique nonnegative integer $n$, the ``classical
dimension'' of the triple, such that~\eqn{Cesaro1} holds.
This is known to be valid~\cite{Odysseus} both for discrete
and continuous spectra. In~\cite{Odysseus} it is also
indicated what the subsequent coefficients are (in the
Ces\`{a}ro or average sense) and how to compute them.

\item{\it Regularity\/}:
There is no need to modify the second ``axiom": the
commutative case follows from regularity theorems for
Sobolev spaces, just as in the compact case.

\item{\it Reality\/}:
There is no need to modify the third ``axiom". One takes for
$C$ the usual charge conjugation operator for spinors. One
finds
\be
Cf^*\star C^{-1}\Psi = \Psi\star f,
\label{wondrouseq}
\ee
so indeed we have a commuting representation.

\item{\it First order\/}:
There is no need to modify the fourth ``axiom".

\item{\it Finiteness\/}:
First of all, the algebra $\cal S$ is a pre-$C^*$-algebra, both for the
ordinary and the Moyal product, and this is very easy to prove. We recall
very briefly the argument from~\cite{Polaris}. Let $(1+f)(1+g)=1$, with
$f\in\cal S$; then $g$ must be in ${\cal O}_0$, and then in $\cal S$. The
analogous property can be checked as well for the star product, using the
sequence space decomposition introduced in~\cite{Phobos}.

In order to deal with the rest of the finiteness requirement (and also for
eventual consideration of the $K$-theory groups), we invoke unitizations
(i.e., compactifications) $\tilde{\A}$ of $\A$. Consideration of the
multiplier algebra ${\cal O}_{M}$ of polynomially bounded smooth functions,
together with all their derivatives, and its subalgebras ${\cal O}_{T}$, in
which the degree of the polynomial bound can go up only by one with each
derivative, ${\cal O}_{c}$, already defined, and ${\cal K}$, in which the
degree of the polynomial bound goes down by one with each derivative,
impose themselves, as suitable compactifications of ${\cal S}(\real^4)$.

Also, for us, the {\it Moyal algebra\/}, denoted
$M({\real}_\theta^4)$ or simply $M_\theta$, is the {\it
maximal\/} unitization of the Schwartz algebra $\cal S$ with
the Moyal product. The precise definition is dealt with in
the appendix, but we bring some material up here. Concerning
the Moyal multiplier algebra there have been misleading
statements in the literature recently as well. We welcome
the occasion to clarify matters.

Already we have asserted that, if $f,g$ are Schwartz,
$f\star_\theta g$ is also a Schwartz function. The product
operation is {\it continuous\/} and therefore the tracial
identity
$$
\int f \star_\theta g = \int fg,
$$
valid for any $\theta$, allows the extension of the Moyal product via
linear space duality~\cite{Phobos}, to large classes of distributions. Thus
the expressions $T\star_\theta f,f\star_\theta T$, with $T$ a tempered
distribution, make sense; moreover, the multiplication is (separately)
continuous on its two variables. Witness to the excellent smoothing
properties of the Moyal operation, these products are always smooth. We say
$T\in M_\theta$ if both $T\star_\theta f$ and $f\star_\theta T$ belong to
$\cal S$, for all $f\in\cal S$.

On the ``size" of $M_\theta$ there have been conflicting claims. It
certainly contains ${\cal K}$ and ${\cal O}_c$ as
subalgebras~\cite{Amalthea}, for all $\theta$. But for instance, the
authors of~\cite{MormonChurch} seem to doubt that it is any bigger, whereas
there have been claims that it includes {\it all\/} of tempered
distributions!  The truth is intermediate: whereas $M_\theta$ is large,
containing in particular many distributions, it does not contain ${\cal
O}_T$, and thus ${\cal O}_M$, the commutative multiplier algebra. The true
net of algebras was described in detail in~\cite{Phobos} and~\cite{Deimos}.
At the end of the appendix we exemplify a family of nice-looking, smooth,
bounded functions in ${\cal O}_T$, to wit, exponentials with quadratic
exponents. For the most part, they belong to $M_\theta$, but there are
exceptions. This will show: a) $M_\theta$ is bigger than ${\cal O}_c$; b)
$M_\theta$ is smaller than ${\cal O}_T$, and thus than ${\cal S}'$; c)
$M_\theta$ depends on $\theta$.

\smallskip

We now postulate that the space of smooth vectors $\bigcap_k
\opname{Dom}(D^k)$ is a finitely generated projective left
$\tilde{\A}$-module, for some appropriate unitization
$\tilde{\A}$. It is clear that the module of spinors foots
the bill.

\item{\it Orientation\/}:
We shall keep the Hochschild condition, but postulated for
elements of $\tilde{\A}$ and
$\tilde{\A}\otimes\tilde{\A}^\circ$. More precisely, we
shall assume that it is possible to represent on $\hil$
elements $a$ of $\tilde{\A}$ such that the commutators
appearing in the Hochschild equation are still bounded and
the same equation~\eqn{Hochschildeq} holds.

It is really easy to see that this condition is verified,
too, in the commutative case. The normalized volume form,
say for $n=4$, is
$$
dx_1 \wedge \cdots \wedge dx_4,
$$
with $x_1:=q_1;\;\dots\;;\;x_4:=p_2$. The corresponding
Hochschild $4$-chain in $\tilde{\cal S}({\real}^4)$ is
obtained by applying the skew\-sym\-metrization operator
$$
c:= -\frac{1}{4!} \sum_{\sigma\in S_4} (-1)^\sigma x_{\sigma(1)} \otimes
x_{\sigma(2)} \otimes x_{\sigma(3)} \otimes x_{\sigma(4)}.
$$
This $n$-chain is patently a Hochschild cycle. We can represent it
on~$\hil$, according to
$$
\pi_D(c):= -\frac{1}{4!} \sum_{\sigma\in S_4} (-1)^\sigma \,[D,x_{\sigma(1)}]
 \cdots
\,[D,x_{\sigma(4)}].
$$
Then $\pi_D(c) = \gamma_5 = \chi$.

For the Moyal product this computation is identical as for
the commutative one! But $c$ in this case must be a
Hochschild 4-boundary, so it cannot serve the essential
purpose of representing the fundamental class of a
noncommutative manifold.

We are at a loss to replace the orientability axiom in the
context of star triples. The formal computation can in
principle always carried out, though ---see the discussion
in next section.

\item{\it Poincar\'{e} duality\/}:
Let us just say that some nondegenerate pairing between the $K$-theory of
$\A$ and its $K$-homology with compact supports must exist. That this
happens in the commutative case follows from de Rham theory.
\end{enumerate}

\section{Noncompact star triples by deformation of $\realb^3$}

Beyond the commutative case, the Moyal one for star triples, and the
noncompact spaces associated to the Connes--Landi and
Connes--Dubois-Violette spheres,
a natural question is whether the new star products give rise to noncompact
noncommutative spin geometries, or at least to star triples in the sense
examined in the last section.

The analysis of the contents of this Pandora's box is still
far off. However, some candidates to become ``the" Dirac
operators present themselves. Among the candidates having a
sporting chance, one discerns at once operators of the form
\be
D_\sigma =-i\left( \sigma_x\{x,.\} + \sigma_y\{y,.\} +
\sigma_w\{w,.\}\right),
\label{hopeful}
\ee
acting on $L^2({\real}^3)\oplus L^2({\real}^3)$, with
$-i(\sigma_j\partial_j)$ substituted for them whenever $x_j$
becomes central. These are actually differential operators,
called sometimes Hamiltonian and sometimes Poissonian
operators. Implicitly, we are using here the standard
Riemannian structure on $\real^3$. It is not quite clear
whether the corresponding Riemannian measure is appropriate
for the needs of field theory.

An immediate, if apparently minor, difficulty presents
itself: the coefficients of the partial derivatives in these
differential operators are no longer constant, but linear
functions of the coordinates. That means that ellipticity of
$D_\sigma$ is spoiled at the origin (and in some cases, at
hyperplanes going through it), whereas the Wodzicki residue
diverges at the same point(s). If we put aside this
``ultraviolet divergence", we can see that postulate 1 for
star triples is fulfilled for $D_\sigma$, in view of the
general properties of elliptic operators~\cite{Odysseus}.

Postulate 2 is inherited from the Moyal case. Postulate 3 can be guaranteed
by construction, just as in the standard compact case. Note
that~\eqn{wondrouseq} will still hold, for the $\ast_{\cal G}$ product. (We
suppress ${\cal G}$ from the notation henceforth.) Postulate 4 can be
satisfied in a similar way. All this means that, formally speaking, it is
perfectly possible to define fermion fields on which our algebras act
componentwise by the star product.

Postulates 5 (abstract first order) and 6 (orientability), represent
equations for $D$ which are not easy to fulfil in general; in
particular~\eqn{Hochschildeq} amounts to a difficult nonlinear equation.
The strategy for fulfilling postulate 5 is transparent: if a tentative $D$
is a quasiderivation, that is:
$$
D(f\ast g) = {\hat D}f\ast g + f\ast Dg,
$$
for some appropriate ${\hat D}$, then $[D,f]_\ast$ fulfils
$$
[D,f]_\ast \ast g = {\hat D}f\ast g,
$$
and thus it commutes with right $\ast$ multiplication. The operators
$D_\sigma$ of~\eqn{hopeful} are first order, in this abstract sense.

On the other hand, we have been generally unable to solve
for~\eqn{Hochschildeq} in a satisfactory way in all cases.
This matters because it seems unlikely to us that these
versions of the enveloping algebras have trivial Hochschild
homology.  We have checked, however, that with the operator
$D_\sigma$ it is possible to find a suitable formal cocycle
for $e(2)$, $iso(1,1)$, and $sb(2,\complex)$. For the
Heisenberg algebra, in order to take into account the
central element {\it and\/} to find the right
dimensionality, we add as advertised the $-i\sigma_3\del_w$
term; we are very close there to the Moyal case. It is not
too difficult to show that, for example in the case $e(2)$
the choice $a_0=(12 w^2y)^{-1}$, $a_1=x,a_2=y$ and
$a_3=x+w$, at the price of an $a_0$ singular on two
hyperplanes, solves the equation:
\be
\varepsilon_{ijk}\, a_0\ast[D_\sigma,a_i]_\ast\ast[D_\sigma,a_j]_\ast
\ast[D_\sigma,a_k]_\ast = 1\ .
\label{hochhope}
\ee
Similarly, other linear combinations for the $x_i$ and a
cubic inverse $a_0$ solve the other cases. In all cases the
presence of a null $\ast$-commutator is crucial in the
calculation. We have been unable to find a solution for the
case of the semisimple algebras.

The different peculiarities of the products must have to do
with Lie algebra cohomology; but at present we can just
speculate. It might be that some enveloping algebras are not
orientable in Connes' sense. This likely ``fact of nature",
as in the Moyal case, would not of course mean the
corresponding new star products are physically
uninteresting.

\section{Conclusions}

As we said in the Introduction, this paper has two main
subjects. On one side there is the presentation of a new
machinery to construct noncommutative spaces; on the other
it deals with the issue of noncompact noncommutative
differential geometries.

The method to construct deformed products we have introduced
is based on the idea that a simpler product on a linear
space can give rise under favourable circumstances, via a
reduction process, to another product on a different
manifold. This idea has tremendous potential. We have
explored the reduction from four to three dimensions of a
linear Poisson bracket corresponding to the Moyal product in
four dimensions. But there is no obstruction in principle to
generalize this machinery to the reduction from an arbitrary
number of dimensions. In this case there will be many more
algebras, with richer structures.

Another generalization can go along the lines of considering
non (quadratic-)linear realizations of the linear Poisson
brackets, or nonlinear brackets altogether. Yet another
generalization can be to replace Lie algebras by some of
their generalizations, such as Lie algebroids. And the list
can go on. What we have done in this paper is to just play
with the simplest instances of these products; but
potentially there is a lot more to come.

The second theme of the paper stresses the issue of
noncompactness of these geometries. Our motivations for this
stress are mainly of physical origin, although they have
bearing on mathematics; this is an important issue, even if
the conclusions can only be tentative at the present stage.
In the occurrence, we were faced with the task of
generalizing the requirements spelled out by Connes for the
compact case. We found this to be a feasible task,
especially for (the reduction of spaces which are)
deformations of commutative spaces. Hence the concept of
star triple, for which the new products provided examples.

It is not clear that in all cases the orientability
requirement, already in trouble for Moyal algebra, is
satisfied for the putative Dirac operators. While one
distinct possibility is that we have not been clever enough
to find the right operator, we should not dismiss the
possibility that orientability, as adapted from compact
triples, is a requirement that may require further
modification for the noncompact case.

A third strand running through the paper is the exploitation of the
``nonperturbative" form of the Moyal product, and the construction of the
multiplier algebra therefrom. Even if the Moyal algebra does not quite
reach the lofty status of ``noncommutative spin manifold", its analysis in
depth pays dividends. In effect, the results of this paper point out to a
kind of universal role of the Heisenberg group in noncommutative harmonic
analysis.

{}From the formal point of view much remains to be done. We
have not dealt with the issue of irreducibility. Nor have we
seriously dealt with the proper measure to use on the new
noncommutative spaces, an important aspect if the aim is to
have field theories on them. The measure is likely to be
connected with the orientability issue, and so the study of
the former may shed some light on the latter.

\subsection*{Acknowledgments}

We thank A.~El~Gradechi, J.~Grabowski, H.~Grosse, R.~Salmoni,
A.~Sitarz, R.~Szabo,
R.~Wulkenhaar and A.~Zampini for helpful discussions, A.~P.~Balachandran
for stressing to us the relevance of the oscillator representation for the
fuzzy sphere, and J.~C.~V\'{a}rilly for carefully reading a preliminary version
of the manuscript and many useful suggestions. The work of F.~L., G.~M.\
and P.~V.\ was supported in part by the {\sl Progetto di Ricerca di
Interesse Nazionale {\em SInteSi 2000}}. J.~M~.G.-B., F.~L.\ and P.~V.\
like to thank the Erwin~Schr\"{o}dinger Institute in Vienna for hospitality
during the course of this research.

\setcounter{section}{0}

\appendix{Moyal product primer}

This appendix recapitulates properties of Moyal multiplication relevant for
our endeavour. Different (nondegenerate) Moyal products on $\real^4$ are in
principle associated with an invertible antisymmetric matrix $\Theta_{ij}$
which, with a change of coordinates, can be expressed in the canonical
form:
\be
\Theta=\left(
\begin{array}{cccc}
0& 0& -\theta_1&0\\0&0&0& -\theta_2\\ \theta_1 & 0 & 0 & 0 \\ 0& \theta_2
&0&0
\end{array}
\right),
\ee
with ${\pm}\theta_i$ the eigenvalues of $\Theta$.  A simple rescaling can then
equate $\theta_1=\theta_2=\theta$; and for simplicity we have assumed the
canonical
Poisson bracket for the final coordinates. In this setting, the Moyal product
$f\star_\theta g$ of two Schwartz functions $f,g$ on $\real^4$ is defined by
\begin{equation}
f \star_\theta g(u) := \int_{\reals^4}\int_{\reals^4}\,
L^\theta(u,v,w)\,f(v)g(w)
\,d\mu^\theta(v)\,d\mu^\theta(w), \label{eq:Moyal-prodint}
\end{equation}
where $u:=(q,p);\;\theta$ is a positive real parameter;
$d\mu^\theta(v):= (\pi\theta)^{-4}\,d\mu(v)$ and the integral kernel
$L$ is given by
\begin{equation}
L^\theta(u,v,w):= \exp\biggl(\frac{2i}{\theta}\bigl(uJv + vJw +
wJu\bigr)\biggr),
\label{eq:Moyal-prod-ker}
\end{equation}
where $J$ denotes the antisymmetric matrix:
\be
J:=\left(\begin{array}{cc} 0 & \id_2 \\ -\id_2 & 0 \end{array}\right),
\label{laJ}
\ee
with $\id_2$ the $2{\times} 2$ identity matrix. Equivalent integral
alternative formulae are
\begin{equation}
f \star_\theta g(u) := (\pi\theta)^{-4}
\int_{\real^4}\int_{\real^4}\,f(u + s)g(u + t)\,e^{2isJt/\theta}\,ds\,dt,
\label{eq:Moyal-prodintbis}
\end{equation}
or
\begin{equation}
f \star_\theta g(u) := (2\pi)^{-4}
\int_{\real^4}\int_{\real^4}\,f(u + \half\theta Js)g(u + t)
\,e^{-ist}\,ds\,dt.
\label{eq:Moyal-prodinttris}
\end{equation}
The popular Moyal series development is an {\it asymptotic\/} expansion
of the previous in powers of~$\theta$~\cite{Nereid}:
\begin{equation}
f \star_\theta g(u) \sim fg(u) + i\frac{\theta}{2}\{f,g\}(u)
+
\sum_{k=2}^\infty\,\biggl(i\frac{\theta}{2}\biggr)^k\frac{1}{k!}
\,D_k(f,g)(u),\quad\mbox{as}\ \theta\to0,
\label{eq:Moyal-prodintcuatris}
\end{equation}
where the $k$-order bidifferential transvection operators
$$
D_k(f,g)(q,p) = \frac{\partial^k f}{\partial q^k}\;\frac{\partial^k g}{\partial
p^k} - {k \choose 1} \frac{\partial^k f}{\partial q^{k-1}\partial p}\;
\frac{\partial^k g}{\partial p^{k-1}\partial q} + \ensuremath \cdots + (-)^k
\frac{\partial^k f}{\partial p^k}\;\frac{\partial^k g}{\partial q^k}
$$
were also introduced by Lie~\cite{SophusII} ---in relation to aspects of
classical invariant theory nowadays all but forgotten.

The development \eqn{eq:Moyal-prodintcuatris} becomes exact
under conditions that are spelled out in~\cite{Nereid}.
Outside them, the integral form~\eqn{eq:Moyal-prodint} or
its siblings should be exclusively used. This is illustrated
by the dramatically different nature of the divisors of zero
for~\eqn{eq:Moyal-prodint} and respectively
for~\eqn{eq:Moyal-prodintcuatris}: the Moyal product of two
functions whose supports do not meet in general is {\it
not\/} zero. Also, ``nonperturbative" properties like
T-duality in the periodic case are almost evident from the
correct, integral expressions.

{}From any of the previous formulae it follows that
$(f\star_\theta g)^* = g^*\star_\theta f^*$, with $f^*$
denoting complex conjugate of $f$. It is not difficult to
verify that, as asserted at the end of Section~5, $f
\star_\theta g$ is also a Schwartz function. The product operation is
{\it continuous\/} on $\mathcal{S}$; we refer the reader to~\cite{Phobos},
for all this and the key tracial identity
$$
\int f \star_\theta g = \int g \star_\theta f = \int fg,
$$
which is valid for any $\theta$.

The last two remarks allow the extension of the Moyal
product to large classes of distributions via linear space
duality~\cite{Phobos}, by the formula
$$
\<T\star f,g> := \<T,f\star g>,
$$
for $T$ a tempered distribution (and similarly for the product with $T$
from the right).

In fact, due to the excellent smoothing properties of the
Moyal product, $T\star f$ is always a smooth function,
although in general {\em not} of the Schwartz class. Now
$M_{L}({\real}_\theta^4)$, the left multiplier algebra, is
defined as the subspace of tempered distributions that give
rise to Schwartz functions when left multiplied by Schwartz
functions; it takes no time to check that
$M_L({\real}_\theta^4)$ is indeed an algebra. The right
multiplier algebra~$M_R({\real}_\theta^4)$ is analogously
defined.

The unital {\it Moyal algebra\/} $M_\theta$ is then defined as $M_\theta:=
M_L({\real}_\theta^4)\cap M_R({\real}_\theta^4)$. The algebras $M_\theta$
for different $\theta$ are all different, but they are related by a simple
dilation. In previous sections we referred to several interesting unital
subalgebras of the Moyal algebra $M_\theta$, and to the fact that ${\cal
O}_M$, the space of polynomially bounded smooth functions, together with
all their derivatives, is not one of them. $M_\theta$ is invariant by
rescaled Fourier transform, to wit
$$
FM_\theta=M_{4/\theta};
$$
therefore ${\cal O}'_M\in M_\theta$; in fact, the bigger ${\cal O}'_T\in
M_\theta$. The dual space of distributions $M'_\theta$ is a dense ideal of
the multiplier algebra $M_\theta$, endowed with the natural locally convex
topology. The $M_\theta$ are normal spaces of distributions, and all their
derivations are inner.

We turn to the analysis of a relevant group of automorphisms of $M_\theta$.
Translations of $\real^4$ and real symplectic $4{\times}4$ matrices
(defined by ${}^tSJS = J$) act on functions respectively by
\be
sf(u):= f(u-s);  \quad Sf(u) := f(S^{-1}u). \label{eq:group-act}
\ee
Let $(s,S)$ denote an element of the inhomogeneous symplectic group
$ISp(4;\real)$, i.e., the semidirect product of the group of translations
and the symplectic group, with group law
$$
(s_1,S_1)(s_2,S_2) = (S_2^{-1}s_1+s_2,S_1S_2).
$$
As an immediate consequence of~\eqn{eq:Moyal-prodint}
and~\eqn{eq:group-act} we have equivariance of the twisted product:
\begin{equation}
(s,S)f\star_\theta (s,S)g = (s,S)(f\star_\theta g).
\label{eq:lin-symplinv}
\end{equation}

We shall verify presently that the $(s,S)$ are {\it inner\/} automorphisms.
This is clear for the $(s,\id)$ transformations, which correspond to
ordinary exponentials $\exp(-isu)$, that is to say, they generate by the
Moyal star product the Weyl algebra. So we concentrate on the $(0,S)$
transformations.

The Lie algebra $sp(4;\real)$ of infinitesimal symplectic transformations
is formed by matrices $L$ such that
$$
JL+{}^tLJ=0.
$$
Linear and quadratic functions double as Hamiltonians respectively for
translations and linear symplectomorphisms; they generate infinitesimal
linear inhomogeneous symplectic transformations. To be precise, note first
that the matrix $B$ is symmetric iff $JB$ is infinitesimally symplectic.
Let then $h_{(b,B)} = \frac{1}{2}\,{}^tuBu + {}^tbu$, where $B$ is
symmetric. Then $(b,L)\mapsto h_{(b,-JL)}$ is an isomorphism of Lie
algebras between $isp(4;\real)$, with an obvious notation, and the Poisson
subalgebra of quadratic-linear functions on $\real^4$; in particular
$\{h_A,h_B\}$ vanishes iff $[JA,JB]=0$.

It is useful to have explicit formulae for Moyal star products in which
one of the factors $h$ is a function of this kind. Using the asymptotic
development of the Moyal product, here exact, one easily
obtains~\cite{TritonAriel}:
\begin{equation}
h \star_\theta f(u) = hf(u) +
\frac{i\theta}{2}\,{}^t(Bu + b)J \opname{grad}f(u) +
\frac{\theta^2}{8}\Tr\bigl[BJ \opname{Hess}f(u) J\bigr].
\label{eq:pract-prodform}
\end{equation}
It follows either from~\eqn{eq:lin-symplinv} or~\eqn{eq:pract-prodform} that
\begin{equation}
[h,f]_{\star_\theta} = i\theta\,\{h,f\},
\label{eq:Moyal-Poissonmagic}
\end{equation}
i.e., the Moyal bracket and the Poisson bracket in the aforementioned case
essentially coincide.

The linear symplectic group $Sp$ is not simply connected, and it is
pertinent to consider its twofold covering, the so called metaplectic group
$Mp$. Just as in the spin representation, to each element $A\in Mp$
corresponds a symplectic matrix $S(A)$, with $S(A)=S(-A)$. The elements of
$Mp$ are realized by unitaries $\Xi_S$, belonging to the multiplier Moyal
algebra $M$, and defined up to a sign, such that
\be
\Xi_S\star f\star\Xi^*_S = Sf,
\label{internaut}
\ee
for all $f$. Explicitly,
\be
\Xi_S(u) = e^{i\alpha}\frac{4}{\sqrt{\det(\id_4+S)}}
\exp\biggl(-{}^tu\,\frac{iJ(\id_4-S)}{\theta(\id_4+S)}\,u\biggr).
\label{internautbis}
\ee
Note that $S\mapsto(\id_4-S)(\id_4+S)^{-1}$ is the Cayley
transform, sending a symplectic matrix into an
infinitesimally symplectic one: the matrix in the exponent
must be symmetric. After many years of undeserved neglect,
this important ``nonlinear" relation between a Lie group and
its Lie algebra has received an authoritative study
in~\cite{KM}. The $\Xi_S$ are elements of the Moyal algebra
$M_\theta$ not belonging to ${\cal O}_c$. The phase factor
in~\eqn{internautbis} reflects the ambiguity inherent
to~\eqn{internaut}, which, as indicated, can be reduced to a
sign, so that
\be
\Xi_S\star\Xi_{S'}={\pm}\Xi_{SS'}.
\label{productlaw}
\ee
The curious reader can directly check this formula with the
help of the method of the stationary
phase~\cite{TritonAriel}.

Of course, the ambiguity can be eliminated completely for
uniparametric subgroups. Equations~\eqn{internautbis}
and~\eqn{productlaw} are the starting point for a
presentation of our algebras in terms of unitaries and
relations. One has $\Xi_{M^{-1}SM}(u) = \Xi_S(Mu)$, where
$M$ is a symplectic matrix (which under some circumstances
can be taken complex). This simplifies the calculation of
$\Xi_S$, as it suffices to carry it out for ``normal forms",
representative of the orbits of the symplectic group under
the adjoint action. Consider
$S(\beta)=\id_4\cos\beta+J\sin\beta\in Sp(4;\real)$. Then
\be
\Xi_{S(\beta)} = \sec^2(\beta/2)
\exp(-{}^tu\frac{i}{\theta}\tan\frac{\beta}{2}u).
\label{unipargroup}
\ee
Now consider, for instance, the product associated to
$su(2)$. All the generators are {\it elliptic\/}, which
means that $\det B>0$, for the associated symmetric matrix.
Therefore
$$
e^{{\pm} itx_j\star} =
\sec^2(t/2)\exp({\pm}\frac{2ix_j}{\theta}\tan\frac{t}{2}).
$$
The product associated to $sl(2,{\real})$ contains hyperbolic elements
$(\det B<0)$ as well; for them, hyperbolic functions replace the
trigonometric ones in the formula. Parabolic elements are considered
in~\cite{TritonAriel}. Relations can be obtained simply by
using~\eqn{productlaw}.

For ``exceptional" symplectic matrices with $\det(\id_4+S)=0$,
formula~\eqn{internautbis} appears not to be well defined. In fact, we only
need to tackle the case $S=-\id$, corresponding to $\beta={\pm}\pi$ in the
example, and the cases
\[
S_1 = \left(\begin{array}{cccc} -1 & & & \\ & 1 & & \\ & & -1 & \\ & & & 1
\end{array}\right);\quad
S_2 = \left(\begin{array}{cccc} 1 & & & \\ & -1 & & \\ & & 1 & \\ & & & -1
\end{array}\right).
\]
One goes to the distributional limit of the previous
expression~\eqn{unipargroup}, obtaining
$$
\Xi_{S({\pm}\pi)}(u) = \mp i\pi^2\theta^2\delta(u);
$$
for the other cases, a factor $\pi\theta\delta^{(2)}(Pu)$,
is similarly obtained, where $P$ projects over the
corresponding two dimensional symplectic subspace. Notice
that $\Xi_{S({\pm}2\pi)}=-1$, in spite of the fact that
$S(2\pi)=1$: the twofold covering is unavoidable. For other
exceptional elements, it is enough~\cite{TritonAriel} to
factor out $S_1$ or $S_2$ or both, and use the product
law~\eqn{productlaw}.

Consider now the hyperbolic uniparametric subgroup
$S(\alpha)=\id_4\cosh\alpha+A\sinh\alpha\in Sp(4;\real)$,
where
\[
A = \left(\begin{array}{cccc} -1 & & & \\ & -1 & & \\ & & 1 & \\ & & & 1
\end{array}\right)\ .
\]
Then
\be
M({\real}_\theta^4)\ni\Xi_{S(\alpha)} = \mbox{sech}^2(\alpha/2)
\exp(-\frac{i\,2pq\,\tanh(\alpha/2)}{\theta}).
\label{unipargroupbis}
\ee
It is plain that the value $e^{-i\,2pq/\theta}$ is never
reached. For good reason: the functions $e^{\mp
i\,2pq/\theta}$ do {\it not\/} belong to the multiplier
algebra $M({\real}_\theta^4)$. Note the dependence on
$\theta$.

More generally, $e^{{\pm} i\,{}^tuGu/\theta}$ belongs to $M({\real}_\theta^4)$
iff $-JG$ is in the range of the Cayley map; matrices with determinant $-1$
cannot be in that range. These are the ``counterexamples" promised in
Section~6. The Wigner transform can be understood as an operator from
${\cal S}'({\real^{2n}})$ to ${\cal L}({\cal S}({\real}^{n}),{\cal
S}'({\real}^{n}))$ by means of Schwartz's kernel theorem. It sends
$M_L({\real}_\theta^{2n})$ isomorphically onto ${\cal L}({\cal
S}({\real}^{n}),{\cal S}({\real}^{n}))$. In addition to this fact, one
proof that $e^{{\pm} i\,{}^tuGu/\theta}$ does not belong to
$M({\real}_\theta^4)$ when $-JG$ is not in the range of the Cayley map uses
only Fourier analysis.

\end{document}